\documentclass[preprint,showpacs,preprintnumbers,amsmath,amssymb]{revtex4-1}
\usepackage{graphicx}
\usepackage{epstopdf}
\usepackage[usenames]{color}
\usepackage{dcolumn}
\usepackage{bm}
\begin{document}
\input{epsf}
\title{Antihydrogen $(\overline{\rm{H}})$ and muonic
antihydrogen $(\overline{\rm{H}}_{\mu})$ formation
in low energy three-charge-particle collisions 
}
\author{Renat A. Sultanov\footnote{rasultanov@stcloudstate.edu\ \ (r.sultanov2@yahoo.com)}
and
D. Guster\footnote{dcguster@stcloudstate.edu}
}

\affiliation{Department of Information Systems \& BCRL
at St. Cloud State University, 
Integrated Science and Engineering Laboratory Facility, St. Cloud, MN 56301-4498}


\date{\today}

\begin{abstract}
A few-body formalism is applied for computation of two different three-charge-particle systems.
The first system is a  collision of a slow antiproton, $\overline{\rm{p}}$, with a positronium atom: Ps$=(e^+e^-)$ $-$ 
a bound state of an electron and a positron. The second problem is a collision of $\overline{\rm{p}}$ with 
a muonic muonium atom, i.e. true muonium $-$ a bound state of two muons one positive and one negative:
Ps$_{\mu}=(\mu^+\mu^-)$. The total cross section of the following two reactions:
$\overline{\rm p}+(e^+e^-) \rightarrow \overline{\rm{H}} + e^-$ and
$\overline{\rm p}+(\mu^+\mu^-) \rightarrow \overline{\rm{H}}_{\mu} + \mu^-$, where 
$\overline{\rm{H}}=(\overline{\rm p}e^+)$ is antihydrogen and
$\overline{\rm{H}}_{\mu}=(\overline{\rm p}\mu^+)$
is a muonic antihydrogen atom, i.e. a bound state of $\overline{\rm{p}}$ and $\mu^+$,
are computed in the framework of a set of coupled two-component
Faddeev-Hahn-type (FH-type) equations. 
%
%
Unlike the original Faddeev approach the FH-type equations are formulated in terms
of only two but relevant components: $\Psi_1$ and $\Psi_2$, of the system's three-body wave function  $\Psi$,
where $\Psi=\Psi_1+\Psi_2$. In order to solve the FH-type equations $\Psi_1$
is expanded in terms of the input channel target eigenfunctions, i.e. in this work
in terms of, for example, the $(\mu^+\mu^-)$ atom eigenfunctions. At the same time
$\Psi_2$ is expanded in terms of the output channel
two-body wave functions, that is in terms of $\overline{\rm{H}}_{\mu}$ atom eigenfunctions.
Additionally, a convenient total angular momentum projection is performed.
This procedure leads to an infinite set of one-dimensional coupled integral-differential 
equations for unknown expansion functions. Since the two-body targets are treated
equally and the accurate asymptotes of $\Psi_1$ and $\Psi_2$ are provided, the 
solution of the FH-type equations avoids the over-completeness problem. Results
for better known low-energy $\mu^-$ transfer reactions from one hydrogen isotope to another
hydrogen isotope in the cycle of muon catalyzed fusion ($\mu$CF) are also computed and presented.

%

\end{abstract}
\maketitle

\section{Introduction}
\label{sec:intro}
To date, non-relativistic quantum-mechanical Coulomb few-body problems have
a long research history. In fact, the first works dealing with quantum few-charged particle systems
appeared during the early stages of quantum-mechanics \cite{oppenh27}.
This is because these types of problems pose significant fundamental
theoretical and practical importance in nuclear and atomic-molecular physics.
The few-body Coulomb problem is of considerable importance
in the cycle of $\mu$CF (cold fusion) \cite{nagam03a}, in cases where a muonic few-body system
experiences a strong interplay between Coulomb and nuclear forces involving heavy nuclei, for instance,
the (dt$\mu)^+$ molecular ion. Further, it would be worth mentioning that there are modern
antimatter physics problems, that involve few-body systems, such as
antihydrogen \cite{humber87,charlton90,mitroy94,igarashi94a,mitroy97,kino02}/protonium
\cite{sakimoto02,esry03,macek05,tong06,igarashi2008} formation reactions,
low energy   $\overline{\mbox p}$+H$_2^+$ collisions \cite{sakimoto04},
$\overline{\rm{H}}$+H$_2$ quenching \cite{armour08,my10}, $\overline{\rm{H}}$+H
annihilation reactions \cite{dalg}, and $\overline{{\mbox p}}$+$^{4(3)}$He
antiprotonic helium atom (atomcule) formation \cite{yamazaki02,hayano07} just to name a few.
It follows from the charge conjugation,
parity, and time reversal (CPT) symmetry of quantum electrodynamics that a charged particle 
and its antiparticle should have equal/opposite charges, equal masses, lifetimes, 
and gyromagnetic ratios. Moreover, the CPT symmetry predicts that hydrogen and antihydrogen 
atoms should have identical spectra. New experiments are in progress to test these fundamental 
laws and theories of physics involving antiparticles as well as antimatter in general.
Therefore, future experimentalists
plan to test whether H and $\overline{\rm{H}}$ have such properties. In such
experiments, it would be important to have a certain quantity of $\overline{\rm{H}}$ atoms at low kinetic energies,
ideally at the rest: $T\sim0$ K \cite{andresen2010,gabrielse2011}.
Using this perspective we develop a quantum-mechanical
approach which would be reliable at low and very low collision energies, 
i.e. when the quantum-mechanical few-body dynamics of three Coulomb particles
becomes important. The method is formulated for arbitrary masses of the particles, that is
when the dynamics of lighter and heavier particles are not separated from each other.

The author of the book \cite{nagam03} pointed out that the muonic antihydrogen atom,
$\overline{\rm{H}}_{\mu}$, could be even a better choice to check the CPT law than the usual
antihydrogen atom. This is because the size of this atom is $\sim$ 207 times smaller
than the size of a normal $\overline{\rm{H}}$ atom. Therefore, as mentioned in \cite{nagam03}:
the short range CPT violating interaction with an extremely heavy boson can be easily detected within the system.
This idea appears be extremely interesting, and therefore it would be useful to compute
the formation cross sections and rates of $\overline{\rm{H}}_{\mu}$ at low energy collisions, for example,
from $\sim$1 eV down to $\sim10^{-5}$ eV. Thus, in this work we consider the following three-body reactions
of antihydrogen $\overline{\rm{H}}$ and muonic antihydrogen $\overline{\rm{H}}_{\mu}$ formation:
\begin{equation}\label{eq:Pse}
\overline{\rm p}+(e^+ e^-)_{1s}\rightarrow\overline{\rm{H}} + e^-,
\end{equation}
\begin{equation}\label{eq:Psmu}
\overline{\rm p}+(\mu^+\mu^-)_{1s}\rightarrow\overline{\rm{H}}_{\mu} + \mu^-.
\end{equation}
At such low energies the quantum-mechanical
Coulomb few-body dynamics become important, especially in the case of heavy charge
transfer, i.e. $\mu^+$. Also, it would be quite appropriate to mention that
exotic atomic and antiatomic systems like a true muonium atom, $(\mu^+\mu^-)$,
or a simple muonic hydrogen atom, H$_{\mu}$=(p$^+\mu^-)$, are always of great interest in nuclear, atomic and 
few-body physics \cite{watanabe85a,cohen98,cohen04}. For instance,
recently the authors of works \cite{lebed09,tm2012} have considered an interesting problem: 
the production of $(\mu^+\mu^-)$. This is the smallest pure QED atom with the Bohr radius only $\sim$512 fm.
So far $(\mu^+\mu^-)$ has never been observed.
Next, in the recent works \cite{karr12,tucker11} the proton-radius puzzle \cite{pohl10} was considered
from few-body and muonic physics perspectives.
Another three-charge-particle reaction of $\overline{\rm{H}}_{\mu}$ formation was considered in the works
\cite{cohen04,nagam03} too:
\begin{eqnarray}\label{eq:Psmu2}
\overline{\rm p} + \mbox{Mu} &\rightarrow& \overline{\rm{H}}_{\mu} + e^-.
\end{eqnarray}
Here, Mu is the muonium atom, i.e. a bound state of a positive muon $\mu^+$ and an electron: Mu=$(\mu^+e^-)$.
This is a very interesting and complex example of a heavy charge transfer reaction \cite{cohen04,annual2010}.

In nuclear physics, involving applications related to three-body systems the few-body Faddeev and 
Alt-Grassberger-Sandhas (AGS) equations
\cite{fadd61,fadd93,merk80,sandhas1} are frequently employed. These equations are equal
to the Schr\H{o}dinger equation, but formulated for the three-body wave function components and therefore 
have the correct physical asymptotes. However, in the case of
three-charged particle systems the kernels of the original
integral Faddeev equations in momentum space lose their compactness due to
Coulomb long-range interactions \cite{fadd93}. This limitation has become the most serious obstacle in the
practical application of the original Faddeev equation to few-body systems with pure Coulomb interactions.
Therefore, on one hand the Faddeev and AGS equations are the most rigorous attempt to provide 
a basis for detailed few-body numerical computations, but
on the other hand they have not been used much to date because they 
have been regarded as too complex to solve when used in Coulomb scattering problems. This limitation has led to 
various alternative methods. Among the most popular is a well known method
based on the Born-Oppenheimer adiabatic model \cite{melezhik}    
and improved adiabatic approximation \cite{cohen91}. The approach has been applied to
many systems in atomic and $\mu$-atomic physics for over many years. 
Very accurate variational calculations have been applied to selected three-body Coulomb systems in
muon catalyzed fusion cycle \cite{kami93}, and in $\overline{\mbox{H}}$ formation reactions, see for instance \cite{humber87}.
Here it would also be useful to mention important Coulomb few-body calculations based on the
adiabatic hyper-spherical method \cite{igarashi94a,igarashi94},
coordinate-space Faddeev equation approach in three dimensions \cite{hu93} and
within a hyper-spherical function expansion formalism \cite{kvits95}.                
There are also newer developments in the field we would also like
to cite \cite{jmr05,matv09,armour12}.

In the current work however
we apply a different {\it few-body} approach based on a set of coupled two-component
FH-type equation formalism \cite{hahn72,my99b,my03}.
The next section represents the notation pertinent to the three-charged-particle
system (123) shown in Fig.\ 1, the original equations, boundary conditions, detailed derivation of the set of
coupled one-dimensional integral-differential equations
suitable for a numerical computation and the numerical computational approach developed in this work.
Sec. III includes new results, conclusions, and Sec. IV includes Appendix. 
The atomic units, i.e. $e = \hbar = m_e =1$, are used in the case of the $e^{+}$ transfer reaction, and
the muonic units, i.e. $e = \hbar = m_\mu =1$, are used in the case of the $\mu^+/\mu^-$ transfer reactions,
where $m_{\mu}=206.769\ m_e$ is the mass of the muon.

\section{Few-body treatment}
\label{sec:fbd1}
In the case of three charged particles (123), two positive and one negative, only two asymptotic
configurations are possible below the breakup threshold. This situation is shown in Fig.\ 2, for instance, for the 
$(\overline{{\mbox p}}\  \mu^- \mu^+)$ system. 
It suggests to write down a set of two
coupled equations for Faddeev-type components of the system's wave function \cite{hahn72}.
These equations are commonly called Faddeev-Hahn-type (FH-type) equations \cite{my99b}.
In this work we shall consider a method based on an integral-differential
equation approach \cite{my03} applied to Coulomb three-body systems.
To solve these equations, a modified close coupling method is applied.
This procedure leads to an expansion of the three-body system wave function components
into eigenfunctions of the subsystem Hamiltonians, providing
an infinite set of one-dimensional integral-differential equations \cite{my03}.
Within this formalism the asymptotic of the full three-body wave function contains two 
parts corresponding to two open channels.
In this work
we consider different Coulomb three-body systems with arbitrary masses, i.e.
the masses of the charged particles are taken as they are. We do not apply any type of adiabatic approximations,
when the dynamics of heavy and light parts of the system
are separated.Therefore, this dynamical method works for both a light charge,
$e^+$, transfer reaction at ultra-low energies and for heavy charge transfer, $\mu^+$, as well.

\subsection{FH-type equation approach}
\label{subsec:fh1}
Let us define the system of units to be $e=\hbar=m_3=1$ and
denote antiproton $\overline{\rm{p}}$ by 1, a negative muon $\mu^-$  by 2, and a positive muon $\mu^+$ by 3.
Before the three-body breakup threshold two cluster asymptotic configurations
are possible in the three-body system, i.e. (23)$-$1 and
(13)$-$2  being determined by their own Jacobi coordinates $\{\vec r_{j3}, \vec \rho_k\}$ as shown in Figs. 1, 2:
\begin{eqnarray}\label{eq:coord}
\vec r_{j3} &=& \vec r_3 - \vec r_j,\\
\vec \rho_k &=& (\vec r_3 + m_j\vec r_j) / (1 + m_j) - \vec r_k,\ \ (j \not = k=1, 2).
\end{eqnarray}
Here $\vec r_{\xi}$, $m_{\xi}$ are the coordinates and the
masses of the particles $\xi=1, 2, 3$ respectively.
This suggests a Faddeev formulation which uses only two components.
A general procedure to derive such formulations is described in work \cite{hahn72}.
In this approach the three-body wave function is represented as follows:
\begin{equation}\label{eq:psi3b}
|\Psi\rangle =  \Psi_1 (\vec r_{23},\vec \rho_1) + \Psi_2 (\vec r_{13},\vec \rho_2),
\end{equation}
where each Faddeev-type component is determined by its own Jacobi coordinates. Moreover,
$ \Psi_1 (\vec r_{23}, \vec \rho_1)$ is quadratically integrable
over the variable $\vec r_{23}$, and $\Psi_2 (\vec r_{13},\vec
\rho_2)$ over the variable $\vec r_{13}$. To define $|\Psi_l\rangle$, $(l = 1, 2)$ a set of two coupled
Faddeev-Hahn-type equations can be written:
\begin{equation}\label{eq:fh1}
\Big (E-\hat{H}_0-V_{23}(\vec r_{23})  
\Big ) \Psi_1 (\vec r_{23}, \vec \rho_1) =  
\Big (V_{23}(\vec r_{23}) + V_{12}(\vec r_{12}) 
\Big )\Psi_2 (\vec r_{13}, \vec \rho_2),
\end{equation}
\begin{equation}\label{eq:fh2}
\Big (E-\hat{H}_0-V_{13}(\vec r_{13}) 
\Big )\Psi_2 (\vec r_{13}, \vec \rho_2) =  
\Big (V_{13} (\vec r_{13}) + V_{12}(\vec r_{12}) 
\Big )\Psi_1 (\vec r_{23}, \vec \rho_1).
\end{equation}
Here, $\hat{H}_0$ is the kinetic energy operator of the
three-particle system, $V_{ij} (r_{ij})$
are paired interaction potentials $(i \not= j = 1,2,3)$,
$E$ is the total energy.
The constructed equations satisfy the Schr\H{o}dinger
equation exactly. For the energies below the three-body
break-up threshold they exhibit the same advantages as the
Faddeev equations \cite{fadd93}, 
because they are formulated for the wave
function components with the correct physical asymptotes.
To solve the  equations a close-coupling
method is applied, which leads to an expansion
of the system's wave function components into
eigenfunctions of the subsystem (target) Hamiltonians
providing with a set of one-dimensional integral-differential equations
after the partial-wave projection. A further advantage of the Faddeev-type
method is the fact that the Faddeev-components are smoother functions of
the coordinates than the total wave function.
Also, the Faddeev decomposition avoids overcompleteness problems, because
two-body subsystems are treated in an equivalent way, and the correct
asymptotes are guaranteed. Next, based on
Merkuriev \cite{merk80,fadd93} the three-charge-particle scattering wave function and its all three components
should have the following general asymptotic form:
\begin{eqnarray}\label{eq:asymp}
\Psi_k(\vec r_{j3}, \vec \rho_{k})
&\mathop{\mbox{\large$\sim$}}\limits_{\rho_k \rightarrow + \infty}&
e^{ik_1z} \varphi_1(\vec r_{j3})\delta_{k1} +
\sum_n^{\infty} A^{el/ex}_n(\Omega_{\rho_{k}})
\frac{e^{ik_n\rho_{k}}}{\rho_k}
\varphi_n(\vec r_{j3})+\nonumber \\ &&
\sum_m^{\infty} A^{tr}_m(\Omega_{\rho_{j}})
\frac{e^{ik'_m\rho_{j}}}{\rho_j}\varphi_m(\vec r_{k3})
+ B(\Omega_5)\frac{e^{i(\sqrt{E}\rho + W^c(\rho, E)}}{\rho^{5/2}}.
\end{eqnarray}
Here, $e^{ik_1z} \varphi_1(\vec r_{j3})$ is the incident wave,
$\varphi_n(\vec r_{j3})$ the $n-$th bound-state wave function
of the pair $(j3)$, $k_n = (E-\varepsilon_n)^{1/2}$, $\varepsilon_n$
is the binding energy of the $(j3)$,
$A^{el/ex}(\Omega_{\rho_{k}})$, $A^{tr}(\Omega_{\rho_{j}})$
and $B(\Omega_5)$ are
amplitudes of elastic/inelastic, transfer and breakup channels
respectively, ${\rho_6}=(\rho, \Omega_5)$ is the three-body
hyperradius and $W^c(\rho, E)$ is the three-body Coulomb phase \cite{merk80}.
For lower energy collisions when
$E < E_{thr}$, where $E_{thr}$ is the three-body break-up threshold,
the expression (\ref{eq:asymp}) becomes simpler, i.e. without the last term:
$B(\Omega_5) = 0$.

Therefore, in the current work for low energy collisions each Faddeev type component corresponds to only one determined
channel. For example, for the elastic and for the charge transfer channel we have:
\begin{eqnarray}\label{eq:tr}
\Psi_1(\vec r_{23}, \vec \rho_1)
&\mathop{\mbox{\large$\sim$}}\limits_{\rho_1 \rightarrow + \infty}&
e^{ik_1z}\varphi_1(\vec r_{23})+
\sum_n^{\infty}A_n^{el/ex}(\Omega_{\rho_1})\frac{e^{ik_n\rho_1}}{\rho_1}
\varphi_n(\vec r_{23}),\\
\Psi_2(\vec r_{13}, \vec \rho_{2})
&\mathop{\mbox{\large$\sim$}}\limits_{\rho_2 \rightarrow + \infty}&
\sum_m^{\infty}A_m^{tr}(\Omega_{\rho_2})\frac{e^{ik'_m\rho_2}}{\rho_2}
\varphi_m(\vec r_{13}).
\end{eqnarray}
It is easy to see that the asymptotic behavior of the total wave
function (\ref{eq:psi3b}) becomes similar to equation (\ref{eq:asymp}). 

In addition, we would like to point out, that the few-body FH-type equation approach 
(\ref{eq:fh1})-(\ref{eq:fh2}) is  a quite flexible method. For example, let us briefly consider a
muon transfer reaction in the following low energy collision: 
\begin{equation}
\mbox{Li}^{3+} + (\rm{p}^+\mu^-)_{1s} \rightarrow (\mbox{Li}_\mu^{2+})_{1s}+\rm{p}^+. 
\end{equation}
This reaction has a strong, pure Coulomb interaction in the output channel.
This circumstance can be taken into account by adding
a distortion potential into the FH equations (\ref{eq:fh1})-(\ref{eq:fh2}),
i.e. $U(\rho_2)=(Z_1-1)Z_2/\rho_2$, where $Z_1$ is
the charge of Li$^{3+}$ and $Z_2$(=1) is the charge of the hydrogen isotope:
\begin{equation}\label{eq:Li1}
\Big(E - \hat{h}_{23}(\vec r_{23}) - \hat{T}_1(\vec\rho_1)\Big)\Psi_1 (\vec r_{23}, \vec \rho_1)=     
\Big (V_{23}(\vec r_{23}) + V_{12}(\vec r_{12}) - \frac{(Z_1-1)Z_2}{\rho_2}\Big)
\Psi^C_2 (\vec r_{13}, \vec \rho_2),
\end{equation}
\begin{equation}\label{eq:Li2}
\Big(E - \hat{h}_{13}(\vec r_{13}) - \hat{T}_2(\vec\rho_2) - 
\frac{(Z_1-1)Z_2}{\rho_2}\Big)
\Psi^C_2 (\vec r_{13}, \vec \rho_2)=
\Big (V_{13} (\vec r_{13}) + V_{12}(\vec r_{12})\Big) \Psi_1 (\vec r_{23}, \vec \rho_1).
\end{equation}
The two coupled equations satisfy the Schr\H{o}dinger equation exactly, i.e.
when the Eqs. (\ref{eq:Li1})-(\ref{eq:Li2}) are added to each other the two distortion
potential terms vanish. Also, in these equations we identify the target hamiltonians
$\hat{h}_{23}(\vec r_{23})$ and $\hat{h}_{13}(\vec r_{13})$, as well as the kinetic energy operators, i.e.
$\hat{T}_1(\vec\rho_1)$ and $\hat{T}_2(\vec\rho_2)$. One can see, that by converting the differential
Eq. (\ref{eq:Li2}) into an integral equation we can obtain a Coulomb Green function over the Jacobi
coordinate $\rho_2$ in the output channel. The Green function provides the physically correct
Coulomb asymptotic for the component $|\Psi^C_2\rangle$.
The component $|\Psi_1\rangle$ carries the asymptotic behavior for the elastic and inelastic channels and
the component $|\Psi^C_2\rangle$ carries the Coulomb asymptotic behavior in the transfer channel, that is:
\begin{eqnarray}
\Psi_1(\vec r_{23}, \vec \rho_1)\
&\mathop{\mbox{\large$\sim$}}\limits_{\rho_1 \rightarrow + \infty}&\
e^{ik^{(1)}_1z}\varphi_1(\vec r_{23})\ + \
\sum_n A_n^{\mbox{\scriptsize{el/in}}}(\Omega_{\rho_1})
\frac{e^{ik_n^{(1)}\rho_1}}{\rho_1}\varphi_n(\vec r_{23}), \\
\Psi^C_2(\vec r_{13}, \vec \rho_{2})
&\mathop{\mbox{\large$\sim$}}\limits_{\rho_2 \rightarrow + \infty}&
\sum_{ml} A_{ml}^{\mbox{\scriptsize{tr}}}(\Omega_{\rho_2})
\frac{e^{i(k^{(2)}_m\rho_2 - \pi l /2 + \tau_l - \eta / 2 k^{(2)}_m \ln2k^{(2)}_m\rho_2)}}{\rho_2}
\varphi_m(\vec r_{13}). 
\label{eq:coulomb1}
\end{eqnarray}
Here, $e^{ik^{(1)}_1z} \varphi_1(\vec r_{23})$ is the incident wave,
$\varphi_n(\vec r_{j3})$ the $n$-th excited bound-state wave function
of the pair $(j3)$, $k_n^{(i)} = \sqrt{2M_{i}(E- E_n^{(j)})}$,
with $M_i^{-1}= m_i^{-1} + (1 + m_j)^{-1}\ $, where $m_i$ represents the masses of the heavy
particles $p$ and Li$^{3+}$. Here $E_n^{(j)}$ is the binding energy of  $(j3)$, $i\ne j=1,2$,
$A^{\mbox{\scriptsize{el/in}}}(\Omega_{\rho_{1}})$ and $A^{\mbox{\scriptsize{tr}}}(\Omega_{\rho_{2}})$
are the scattering amplitudes in  the elastic/inelastic and transfer
channels. The Coulomb parameters in the second transfer channel are:
$\tau_l = \makebox{arg} \Gamma(l+1+i\eta/2k^{(2)}_m)$ and $\eta = 2M_2(Z_1-1)/k^{(2)}_n$.
One can see, that this approach simplifies  the solution procedure and provides the correct asymptotic
behavior for the solution below the three-body breakup threshold. Further,
the few-body method has been successfully
applied to different three-body muon transfer reactions \cite{my99b}.

\subsection{Obtaining an infinite set of coupled integral-differential FH-type equations}
\label{subsec:fh12}
Now, let us present the equations (\ref{eq:fh1})-(\ref{eq:fh2}) in terms of the adopted notation
\begin{eqnarray}\label{eq:fadd8}
\Big( E + \frac{1}{2 M_k} \triangle_{\vec \rho_k}  +
\frac{1}{2 \mu_{j}} \triangle_{\vec r_{j3}} -
V_{j3}\Big) \Psi_i (\vec r_{j3}, \vec \rho_k) =
\Big( V_{j3} + V_{jk}) \Psi_{i'}(\vec r_{k3}, \vec \rho_j\Big),
\end{eqnarray}
here $i \not= i' = 1, 2$, $M_k^{-1}= m_k^{-1} + (1 + m_j)^{-1}\,$ and  $\mu_j^{-1} = 1 + m_j^{-1}.$
In order to separate angular variables, the wave function components
$\Psi_i$ are expanded over bipolar harmonics:
\begin{equation}\label{eq:bipolar}
\left \{ Y_{\lambda}(\hat \rho) \otimes Y_l(\hat r) \right \}_{LM} =
\sum_{\mu m}C_{\lambda \mu lm}^{LM}Y_{\lambda \mu}(\hat \rho)Y_{lm}(\hat r),
\end{equation}
where $\hat \rho$ and $\hat r$ are angular coordinates of vectors $\vec \rho$
and $\vec r$; $C_{\lambda \mu lm}^{LM}$ are Clebsh-Gordon coefficients;
$Y_{lm}$ are spherical functions \cite{varshal}. 
The configuration triangle of the particles (123) is presented on the Fig. 1 together with the Jacobi
coordinates $\{\vec r_{23}, \vec \rho_1\}$ and $\{\vec r_{13}, \vec \rho_2\}$ and angles between
them. The centre-off-mass of the whole three-body system is designated as $O$.
The centre-off-masses of the two-body subsystems (23) and (13) are $O_1$ and $O_2$ respectively.
Substituting the following expansion:
\begin{equation}\label{eq:lexpan}
\Psi_i(\vec r_{j3}, \vec \rho_{k}) = \sum_{LM\lambda l}
\Phi_{LM\lambda l}^i
(\rho_k, r_{j3}) \left \{ Y_{\lambda}(\hat \rho_k) \otimes
Y_l(\hat r_{j3}) \right \}_{LM}
\end{equation}
into (\ref{eq:fadd8}), multiplying this by the appropriate biharmonic 
functions and integrating over the corresponding angular coordinates of the vectors
$\vec r_{j3}$ and $\vec \rho_k$, we obtain a set of equations which for the
case of the central potentials has the form:
\begin{eqnarray}\label{eq:part}
\Big(E + \frac{1}{2M_k\rho_k^2}\Big\{\frac{\partial}
{\partial \rho_k}(\rho_k^2
\frac{\partial}{\partial \rho_k}) - \lambda (\lambda + 1)\Big\}+
\frac{1}{2\mu_j r_{j3}^2}\Big\{\frac{\partial}
{\partial r_{j3}} (r_{j3}^2\frac{\partial}
{\partial r_{j3}})-
l(l+1)\Big\} - \nonumber \\
V_{j3}\Big)
\Phi_{LM\lambda l}^i(\rho_k, r_{j3}) = \int d\hat \rho_k
\int d\hat r_{j3} \sum_{\lambda' l'}
W_{\lambda l\lambda ' l'}^{(ii')LM} \Phi_{LM\lambda ' l'}^{i'}
(\rho_j, r_{k3}),
\end{eqnarray}
where the following notation has been introduced:
\begin{eqnarray}\label{eq:effpot}
W_{\lambda l \lambda ' l'}^{(ii')LM} =
\left \{ Y_{\lambda}(\hat \rho_k) \otimes Y_l(\hat r_{j3}) \right \}_{LM}^*
\Big(V_{j3} + V_{jk}\Big)\left \{ Y_{\lambda '}(\hat \rho_j) \otimes Y_{l '}(\hat r_{k3})
\right \}_{LM}.
\end{eqnarray}
To progress from (\ref{eq:part}) to one-dimensional equations, we apply  
a modified close coupling method, which consists of expanding 
each component of the wave function $\Psi_i(\vec r_{j3}, \vec \rho_k)$
over the Hamiltonian eigenfunctions of subsystems:
\begin{equation}
\hat h_{j3} = - \frac{1}{2\mu_j}\nabla^2_{\vec r_{j3}} + V_{j3}(\vec r_{j3}).   
\end{equation}
Thus, following expansions can be applied:
\begin{equation}\label{eq:expan}
\Phi_{LM\lambda l}^i(\rho_k, r_{j3}) = \frac{1}{\rho_k} \sum_n
f_{nl\lambda}^{(i)LM}(\rho_k) R_{nl}^{(i)}(r_{j3}),
\end{equation}
where functions $R_{nl}^i(r_{j3})$ are defined by the following equation:
\begin{equation}\label{eq:eqrnl}
\Big(E_n^i + \frac{1}{2\mu_j r_{j3}^2}
\Big\{\frac{\partial}{\partial r_{j3}} ( r_{j3}^2 \frac{\partial}
{\partial r_{j3}} ) - l(l+1)\Big\} - V_{j3}\Big)R_{nl}^i(r_{j3})= 0.
\end{equation}
Substituting Eq. (\ref{eq:expan}) into (\ref{eq:part}), multiplying by the corresponding functions 
$R_{nl}^i(r_{j3})$ and integrating over $r_{j3}^2dr_{j3}$ yields a set of integral-differential 
equations for the unknown functions $f_{nl\lambda }^i( \rho_k )$:
\begin{eqnarray}
2M_k(E - E_n^i)f_{\alpha}^i(\rho_k)+\Big(\frac{\partial^2}
{\partial \rho_k^2} - \frac{\lambda (\lambda + 1)}{\rho_k^2}  
\Big)f_{\alpha}^i(\rho_k) &=&  \nonumber \\
2M_k \sum_{\alpha}\int_{0}^{\infty} dr_{j3} r_{j3}^2 
\int d\hat r_{j3} \int d\hat \rho_k
\frac{\rho_k}{\rho_j} Q_{\alpha \alpha'}^{ii'}
f_{\alpha'}^{i'}(\rho_j),
\label{eq:part2}
\end{eqnarray}
where
\begin{equation}
Q_{\alpha \alpha'}^{ii'} = R_{nl}^i(r_{j3}) W_{\lambda l \lambda' l'}^{(ii')LM} R_{n'l'}^{i'}(r_{k3}).
\end{equation}
For brevity one can denote $\alpha \equiv nl\lambda$
$(\alpha^\prime \equiv n^\prime l^\prime \lambda^\prime)$,
and omit $LM$ because all functions have to be the same. The functions $f_{\alpha}^i(\rho_k)$ 
depend on the scalar argument, but this set is still not one-dimensional,
as formulas in different frames of the Jacobi coordinates:  
\begin{eqnarray}
\vec \rho_j = \vec r_{j3} - \beta_k \vec r_{k3},\ \ 
\vec r_{j3} = \frac{1}{\gamma}(\beta_k \vec \rho_k + \vec \rho_j),\ \ 
\vec r_{jk} = \frac{1}{\gamma}(\sigma_j \vec \rho_j - \sigma_k \vec \rho_k),
\end{eqnarray}
with the following mass coefficients:
\begin{eqnarray}
\beta_k = m_k / (1 + m_k),\  \
\sigma_k = 1 - \beta_k,\ \
\gamma = 1 - \beta_k \beta_j\ \ (j\not = k = 1, 2),
\end{eqnarray}
clearly demonstrate that the modulus of $\vec \rho_j$ depends on two vectors, over
which integration on the right-hand sides is accomplished: $\vec \rho_j = \gamma \vec r_{j3} - \beta_k \vec \rho_k$.
Therefore, to obtain one-dimensional integral-differential equations,
corresponding to equations (\ref{eq:part2}),
we will proceed with the integration over variables
$\{\vec \rho_j,\hat
\rho_k\}$, rather than $\{\vec r_{j3},\hat \rho_k\}$. The Jacobian of this transformation is $\gamma^{-3}$.
Thus, we arrive at a set of one-dimensional integral-differential equations:
\begin{eqnarray}
2M_k(E - E_n^i)f_{\alpha}^i(\rho_k) + \Big(\frac{\partial^2}
{\partial \rho_k^2} -
\frac{\lambda (\lambda + 1)}{\rho_k^2}   
\Big) f_{\alpha}^i(\rho_k)= 
\frac{M_k}{\gamma^{-3}}\sum_{\alpha'}
\int_{0}^{\infty} d \rho_jS_{\alpha \alpha'}^{ii'}(\rho_j, \rho_k)f_{\alpha'}^{i'}
(\rho_j),\ \ \ \ 
\label{eq:part4}
\end{eqnarray}
where functions $S_{\alpha \alpha'}^{ii'}(\rho_j, \rho_k)$
are defined as follows:
\begin{eqnarray}
S_{\alpha \alpha'}^{ii'}(\rho_j, \rho_k) = 2\rho_j \rho_k
\int d \hat \rho_j \int d \hat \rho_k R_{nl}^i(r_{j3})
\left \{ Y_{\lambda}(\hat \rho_k) \otimes Y_l(\hat r_{j3}) \right \}_{LM}^*
\Big(V_{j3} + V_{jk}\Big)\nonumber \\   
\left \{ Y_{\lambda^\prime}(\hat \rho_j) \otimes Y_{l'}(\hat r_{k3})
\right \}_{LM} R_{n'l'}^{i'} (r_{k3})\;.
\label{eq:ss}
\end{eqnarray}

One can show (see Appendix, Sect. IV) that fourfold
multiple integration in equations (\ref{eq:ss}) leads to 
a one-dimensional integral and the expression (\ref{eq:ss}) could be
determined for any orbital momentum value $L$:
\begin{eqnarray}\label{eq:final}
S_{\alpha \alpha'}^{ii'}(\rho_j, \rho_k)=
\frac{4 \pi}{2L+1}
[(2\lambda + 1)(2\lambda^{\prime} + 1)]^{ \frac{1}{2} }
\rho_j \rho_k \int_{0}^{\pi}
d \omega \sin\omega R_{nl}^i(r_{j3})
\Big(V_{j3}(r_{j3}) +\; \nonumber \\
V_{jk}(r_{jk})\Big)  
R_{n'l'}^{i'}(r_{k3})
\sum_{mm'} D_{mm'}^L(0, \omega, 0)
C_{\lambda 0lm}^{Lm}
C_{\lambda' 0l'm'}^{Lm'} Y_{lm}(\nu_j, \pi) Y^*_{l'm'}(\nu_k, \pi)\;,
\end{eqnarray}
where $D_{mm'}^L(0, \omega, 0)$ are Wigner functions,
$\omega$ is the angle between $\vec \rho_j$ and $\vec \rho_k$,
$\nu_j$ is the angle between $\vec r_{k3}$ and $\vec \rho_j$,
$\nu_k$ is the angle between $\vec r_{j3}$ and $\vec \rho_k$ (see the Fig. 1).
Finally, we obtain an infinite set of coupled integral-differential
equations for the unknown functions $f_{\alpha}^1(\rho_1)$ and
$f_{\alpha^\prime}^2(\rho_2)$ \cite{my03}:
\begin{eqnarray}
\Big( (k^i_n)^2 + \frac{\partial^2}
{\partial \rho_i^2} -
\frac{\lambda (\lambda + 1)}{\rho_i^2}
\Big) f_{\alpha}^i(\rho_i) = g \sum_{\alpha'} 
\sqrt{\frac{(2\lambda + 1)(2\lambda^{\prime} + 1)}  
{(2L+1)}}
\int_{0}^{\infty} d \rho_{i'}
f_{\alpha^\prime}^{i'}(\rho_{i'})
\nonumber \\
\int_{0}^{\pi}
d \omega \sin\omega
R_{nl}^i(r_{i'3})
\Big(V_{i'3}(r_{i'3}) + V_{ii'}(r_{ii'})\Big)R_{n'l'}^{i'}(r_{i3})
\rho_{i'} \rho_i\nonumber \\
\sum_{mm'} D_{mm'}^L(0, \omega, 0)
C_{\lambda 0lm}^{Lm} C_{\lambda' 0l'm'}^{Lm'}
Y_{lm}(\nu_i, \pi)Y^{*}_{l'm'}(\nu_{i'}, \pi).
\label{eq:fh7}                                                                                                        
\end{eqnarray}
For the sake of simplicity $\alpha \equiv (nl\lambda)$ are quantum numbers of a
three-body state and $L$ is the total angular momentum of the
three-body system, $g=4\pi M_i/\gamma^{3}$,
$k^i_n = \sqrt{2M_i(E-E_n^{i'})}$, where $E_n^{i'}$ is the binding energy of the subsystem $(i^{\prime}3)$,
$M_1=m_1(m_2+m_3)/(m_1+m_2+m_3)$ and
$M_2=m_2(m_1+m_3)/(m_1+m_2+m_3)$ are the reduced masses,
$\gamma=1-m_im_{i'}/((m_i+1)(m_{i'}+1))$,
$D_{mm'}^L(0, \omega, 0)$ the Wigner functions,
$C_{\lambda 0lm}^{Lm}$ the Clebsh-Gordon coefficients,
$Y_{lm}$ are the spherical functions,
$\omega$ is the angle between the Jacobi coordinates
$\vec \rho_i$ and $\vec \rho_{i'}$, $\nu_i$ is the angle between 
$\vec r_{i'3}$ and $\vec \rho_i$, $\nu_{i'}$ is the angle
between $\vec r_{i3}$ and $\vec \rho_{i'}$. One can show that:
$\sin\nu_i  = (\rho_k r_{kj}) / \gamma \sin \omega$, and 
$\cos\nu_i = (\beta \rho_i + \rho_k \cos\omega) / (\gamma r_{kj})$.


\subsection{Boundary conditions, cross sections and numerical implementation}
\label{subsec:cs}
To find a unique solution to Eqs. (\ref{eq:fh7})
appropriate boundary conditions depending on the specific physical situation
need to be considered. First we impose: 
\begin{equation}\label{eq:bound0}
f_{nl}^{(i)}(0) \mathop{\mbox{\large$\sim$}}0.
\end{equation}
Next, for the three-body charge-transfer problems we apply the well known ${\bf K}-$matrix formalism.
This method has already been applied for solution of three-body problems
in the framework of the coordinate space Faddeev equations \cite{kvits95}.
For the present scattering  problem with $i +(j3)$ as
the initial state, in the asymptotic region, it takes two solutions to
Eq.(\ref{eq:fh7}) to satisfy the following boundary conditions:
\begin{eqnarray}
\left\{
\begin{array}{l}
f_{1s}^{(i)}(\rho_i)
\mathop{\mbox{\large$\sim$}}\limits_{\rho_1 \rightarrow + \infty}
\sin(k^{(i)}_1\rho_i) + K_{ii}\cos(k^{(i)}_1\rho_i)\; 
\vspace{1mm}\\
f_{1s}^{(j)}(\rho_j)
\mathop{\mbox{\large$\sim$}}\limits_{\rho_j \rightarrow + \infty}
\sqrt{v_i / v_j}K_{ij}
\cos(k^{(j)}_1\rho_j)\;,\\
\end{array}\right.
\label{eq:fh5}
\end{eqnarray}
where $\it K_{ij}$ are the appropriate coefficients, 
%
%
%
and $v_i$ ($i=1,2$) is a velocity in channel $i$.
%
With the following change of variables in Eq. (\ref{eq:fh7}):       
\begin{equation}\label{eq:replace}
{\sf f}_{1s}^{(i)}(\rho_i)=f_{1s}^{(i)}(\rho_i)-\sin(k^{(i)}_{1}\rho_i), 
%
%
\end{equation}
(i=1, 2) we get two sets of inhomogeneous equations which are
solved numerically. The coefficients $K_{ij}$ can be obtained from a
numerical solution of the FH-type equations. The cross sections are given by
the following expression:
\begin{eqnarray}\label{eq:crosssec}
\sigma_{ij} =
 \frac{4\pi}{k_1^{(i)2}}\left |\frac{{\bf K}}
{1 - i{\bf K}}\right |^2 = 
\frac{4\pi}{k_1^{(i)2}}\frac{\delta_{ij}D^2 + {\it K}_{ij}^2}
{(D - 1)^2 + ({\it K}_{11} + {\it K}_{22})^2},
\end{eqnarray}
where ($i,j=1,2$) refer to the two channels and
$D = K_{11} K_{22} - K_{12} K_{21}$. Also, from the quantum-mechanical unitarity principle one can derive that
the scattering matrix $\bf K=$
$\begin{pmatrix} K_{11} & K_{12} \\ K_{21} & K_{22} \end{pmatrix}$
has the following important feature:
\begin{equation}
K_{12}=K_{21}.
\label{eq:unitarity}
\end{equation}
In this work the relationship (\ref{eq:unitarity}) is checked for all considered collision
energies in both antihydrogen cases, i.e. in $\overline{\rm p}+(e^+e^-)$ and in
$\overline{\rm p}+(\mu^+\mu^-)$, and in the case of the muon transfer reactions.


As stated in Sec.\ref{subsec:fh1} the solution of the Eqs. (\ref{eq:fh1})-(\ref{eq:fh2}) involving
both components $\Psi_{1(2)}$ required that
we apply the expansions (\ref{eq:lexpan}) and (\ref{eq:expan}) over the angle and the distance
variables respectively. However, to obtain  a numerical solution for
the set of coupled Eqs. (\ref{eq:fh7}) we only include the -s and -p waves in the expansion (\ref{eq:lexpan}) and limit
$n$ up to 2 in the Eq. (\ref{eq:expan}). As a result we arrive at a truncated set of six coupled integral-differential equations,
since in $\Psi_{1(2)}$ only 1s, 2s and 2p target two-body atomic wave-functions are included.
This method represents a
modified version of the close coupling approximation with six expansion functions.
The set of truncated integral-differential Eqs. (\ref{eq:fh7})  
is solved by a discretization procedure, i.e.
on the right side of the equations the integrals over $\rho_1$ and $\rho_2$
are replaced by sums using the trapezoidal rule \cite{abram} and
the second order partial derivatives on the left side are
discretized using a three-point rule \cite{abram}. By this means
we obtain a set of linear equations for the unknown coefficients $f^{(i)}_{\alpha}(k)$ ($k=1, N_p$):
\begin{equation}\label{eq:numer0}
\left[\ k^{(1)2}_n + D_{ij}^2 -\frac{\lambda (\lambda + 1)}{\rho_{1i}^2}
\right]f_{\alpha}^{(1)}(i)\ -\ {\frac{M_1}{\gamma^3}} \sum_{\alpha'=1}^{N_s} \sum_{j=1}^{N_p}
w_jS_{\alpha \alpha{\prime}}^{(12)}(\rho_{1i}, \rho_{2j})
f_{\alpha'}^{(2)}(j) = 0,\\  
\end{equation}
\begin{equation}\label{eq:numer1}
-\frac{M_2}{\gamma^3} \sum_{\alpha=1}^{N_s} \sum_{j=1}^{N_p}
w_jS_{\alpha \prime \alpha}^{(21)}(\rho_{2i}, \rho_{1j})
f_{\alpha}^{(1)}(j)\ +\ \left[k^{(2)2}_{n'} + D_{ij}^2 -
\frac{\lambda' (\lambda' + 1)}{\rho_{2i}^2}\right]
f_{\alpha'}^{(2)}(i) = B^{21}_{\alpha'}(i).
\end{equation}
Here, coefficients $w_j$ are weights of the integration points $\rho_{1i}$ and $\rho_{2i}$ ($i=1, N_p$), $N_s$ is the
number of quantum states which are taken into account in the expansion (\ref{eq:expan}).
Next, $D_{ij}^2$ is the three-point numerical approximation for the second order differential operator:
$D_{ij}^2f_{\alpha}(i) = 
(f_{\alpha}(i-1)\delta_{i-1,j} - 2f_{\alpha}(i)\delta_{i,j} +
f_{\alpha}(i+1)\delta_{i+1,j}) / \Delta,$   
where $\Delta$ is a step of the grid $\Delta = \rho_{i+1}-\rho_{i}$. The vector $B^{21}_{\alpha'}(i)$ is:
$
B^{(21)}_{\alpha'}(i) = M_2 / \gamma^3  \sum_{j=1}^{N_p}
w_jS_{\alpha' 1s0}^{(21)}(i, j)\sin(k_1\rho_j),
$
and in symbolic-operator notations the set of linear
Eqs. (\ref{eq:numer0})-(\ref{eq:numer1}) has the following form:
$\sum_{\alpha'=1}^{2\times N_s}\sum_{j=1}^{N_p} {\bf A}_{\alpha \alpha'}(i,j) \vec f_{\alpha'}(j) =\vec b_{\alpha}(i).$
The discretized equations are subsequently solved by the Gauss elimination method \cite{forsythe}.
As can be seen from Eqs. (\ref{eq:numer0})-(\ref{eq:numer1})
the matrix $\bf A$ should have a so-called block-structure: there are four main blocks 
in the matrix: two of them related to the differential operators and other two to the integral operators. 
Each of these blocks should have sub-blocks depending on the quantum numbers $\alpha=nl\lambda$ and
$\alpha' = n'l'\lambda'$. The second order differential operators produce three-diagonal sub-matrixes \cite{my03}.
However, there is no need to keep the whole matrix {\bf A} in computer's operating (fast) memory.
The following optimization procedure shows that it would be possible to reduce the memory usage by at least
four times. Indeed, the numerical equations (\ref{eq:numer0})-(\ref{eq:numer1}) can be written in the following way:
$
D_{1}f^1 -  M_1 \gamma^{-3} S^{12} f^2  = 0,\ \mbox{and}\ 
-M_2 \gamma^{-3} S^{21} f^1 + D_{2}f^2 = b.
$
Here, $D_{1}$, $D_{2}$, $S^{12}$ and $S^{21}$ are sub-matrixes of {\bf A}. Now
one can determine that:
$
f^1 = (D_{1})^{-1}M_1/ \gamma^3S^{12}f^2,
$
where $(D_{1})^{-1}$ is reverse matrix of $D_{1}$. Thereby one can obtain a reduced set of linear equations which
are used to perform the calculations:
$\left [ D_2 - M_1 M_2 \gamma^{-6} S^{21} (D_1)^{-1}S^{12}\right]f^2 = b$ \cite{my03}.

To solve the coupled integral-differential equations (\ref{eq:fh7}) one needs to first compute 
the angular integrals Eqs. (\ref{eq:final}). They are independent of energy $E$.
Therefore, one needs to compute them only once and then store them on a computer's hard drive 
(or solid state drive) to support future
computation of other observables, i.e. the charge-transfer cross-sections at different collision energies. 
The sub-integral expressions in (\ref{eq:final}) have a very strong and complicated dependence 
on the Jacobi coordinates $\rho_i$ and $\rho_{i'}$. To calculate $S_{\alpha \alpha'}^{(ii')}(\rho_i, \rho_{i'})$ 
at different values of $\rho_i$ and $\rho_{i'}$ an adaptable algorithm has been applied together with 
the following mathematical substitution:
$
\cos \omega = (x^2 - \beta_i^2\rho_i^2 - \rho_{i'}^2) / (2\beta_i \rho_i\rho_{i'}).
$
The angle dependent part of the equation can be written as the following one-dimensional integral:
\begin{eqnarray}\label{eq:omega}
S_{\alpha \alpha'}^{(ii')}(\rho_i, \rho_{i'}) =
\frac{4\pi}{\beta_i}
\frac{[(2\lambda + 1)(2\lambda^{\prime} + 1)]^{\frac{1}{2}}}{2L+1}
\int_{|\beta_i\rho_i - \rho_{i'}|}^{\beta_i\rho_i + \rho_{i'}}
dx R_{nl}^{(i)}(x)  
\left[-1 + \frac{x}{r_{ii'}(x)}\right] R_{n'l'}^{(i')}(r_{i3}(x))\nonumber \\
\sum_{mm'} D_{mm'}^L(0, \omega(x), 0)C_{\lambda 0lm}^{Lm}
C_{\lambda' 0l'm'}^{Lm'} 
Y_{lm}(\nu_i(x), \pi) Y^{*}_{l'm'}(\nu_{i'}(x), \pi).\ 
\end{eqnarray}
We used a special adaptive
FORTRAN subroutine from the work \cite{berlizov99} in order to carry out the angle integration in (\ref{eq:omega}).
This recursive computer program, QUADREC,
is a better, modified version of the well known program QUANC8 \cite{forsythe}. QUADREC provides
a much higher quality, stable and more precise integration than does QUANC8 \cite{berlizov99}.
Therefore, our results for the three-particle muon transfer reactions presented in
Table I are slightly different from those of our older work \cite{my99b}
where we used the less effective adaptive quadrature code QUANC8 for numerical computation
of the angle integrals (\ref{eq:omega}). The difference between these two results ranges from 
$\sim$9\% to $\sim$15\%. The expression (\ref{eq:omega})
differs from zero only in a narrow strip, i.e. when $\rho_i \approx \rho_{i'}$.
This is because in the considered three-body systems the coefficient $\beta_i$ is approximately equal to one. 
Figures \ref{fig:fig3} and \ref{fig:fig4} show the angle integral 2-dimensional functions (surfaces) (\ref{eq:omega}) 
for the $(\overline{\rm p}\ \mu^- \mu^+)$ system considered herein.
For example, this might involve a few selected atomic/muonic transitions such as:
$S_{1s:1s'}^{(12)}(\rho_1, \rho_2)$ and 
$S_{1s:2s'}^{(12)}(\rho_1, \rho_2)$. 
Only the input channel $\overline{\rm p}+(\mu^-\mu^+)$ of the reaction (\ref{eq:Psmu}) potential surfaces are included.
It is seen, that these surfaces have significantly different geometrical shapes and numerical values.
Therefore, in order to obtain numerically reliable
converged results it is necessary to adequately distribute a very large number 
of discretization points (up to 2200) between 0 and $\sim$90 atomic/muonic units. 
More points are taken near the origin where the interaction
potentials are large and a smaller number of points are needed at larger distances.

\section{Results and conclusions}
\label{sec:results}
In this section we report our computational results.
Five different three-body Coulomb systems have been computed in the framework of
a {\it unique} quantum-mechanical method, i.e. the FH-type equation formalism (\ref{eq:fh1})-(\ref{eq:fh2}). The few-body
approach has been presented in previous sections. In order
to solve the coupled equations (\ref{eq:fh1})-(\ref{eq:fh2})
we use two different and independent sets of target expansion 
functions (\ref{eq:expan}). This is shown in Fig. 2 for the case of the $(\mu^+\mu^-)$ and
$(\overline{\rm{p}}\mu^+)$ targets.
Together with the specific structure of the two coupled FH-type equations in the operator form
this method allows us to avoid the over-completeness problem and the two targets are treated equivalently.
The main goal of this work is to carry out a reliable quantum-mechanical computation of the formation cross 
sections and corresponding rates of the $\overline{\rm{H}}$ and $\overline{\rm{H}}_{\mu}$ atoms at very
low collision energies, i.e. reactions (\ref{eq:Pse}) and (\ref{eq:Psmu}).
However, as a test of the method and our FORTRAN code
we carried out calculations of the three-body cross sections and rates of the $\mu^-$ transfer reactions from d to t:
t + (d$\mu^-)\rightarrow$ (t$\mu^-)$ + d, from p to t: t + (p$\mu^-)\rightarrow$ (t$\mu^-)$ + p,
and from p to d: d + (p$\mu^-) \rightarrow$ (d$\mu^-)$ + p. Here, p, d and t are the hydrogen isotopes:
proton p=$^1$H$^+$, deuterium d=$^2$H$^+$, and tritium t=$^3$H$^+$.
The coupled integral-differential Eqs. (\ref{eq:fh7}) have been solved numerically for the case of
the total angular momentum $L=0$ within the two-level  2$\times$(1s), four-level 2$\times$(1s+2s), 
and six-level 2$\times$(1s+2s+2p) close coupling approximations in Eq. (\ref{eq:expan}). The sign "2$\times$"
indicates that two different sets of expansion functions are applied. Next,
the following boundary conditions (\ref{eq:bound0}), (\ref{eq:fh5}), and (\ref{eq:replace}) have been used.
To compute the charge transfer cross sections the expression (\ref{eq:crosssec}) has been applied.

It would be useful to make a comment about the behaviour of $\sigma_{tr}(\varepsilon_{coll})$
at very low collision energies: $\varepsilon_{coll}\sim0$.
From our calculation we found that the muon transfer cross sections $\sigma_{tr}\rightarrow \infty$ 
as $\varepsilon_{coll}\rightarrow 0$. However, the muon transfer rates, $\lambda_{tr}$, are proportional to the product
$\sigma_{tr}\times v_{c.m.}$ and this trends to a finite value as $v_{c.m.}\rightarrow 0$. Here
$v_{c.m.}= \sqrt{2\varepsilon_{coll}/M_k}$ is a relative center-of-mass 
velocity between the particles in the input channel of the three-body
reactions, and $M_k$ is the reduced mass. To compute the muon transfer rate the following formula is used:
\begin{equation}\label{eq:rate7}
\lambda_{tr} = \sigma_{tr}(\varepsilon_{coll}\rightarrow 0)v_{c.m.}N_0,  
\end{equation}
where $N_0=4.25 \cdot 10^{22}$c.m.$^3$ is the liquid hydrogen density. 

The Coulomb few-body systems mentioned above are of a significant importance in the $\mu$CF cycle
\cite{nagam03a}. In the literature one can find the results of a variety of different calculations of these reactions.
We compare our results with some of this data.
Table I shows cross sections, $\sigma_{tr}$, and corresponding thermal rates, $\lambda_{tr}$,
for all three muon transfer reactions at low collision energies together with some theoretical calculations from older papers
\cite{melezhik,cohen91,kami93,matv09} and with some experimental data from works \cite{13,14,15,16,17,18,19}.
Our FH-type equation results shown in Table I have been computed within the 
$2\times$(1s+2s+2p) approximation in the expansion
(\ref{eq:expan}) for both Faddeev-type components.Therefore, it was actually used with up to six expansion functions.
One can see that our $\sigma_{tr}$ and $\lambda_{tr}$
are in fairly good agreement with previous calculations and experimental data for {\it all three muonic systems}
presented in Table I.
The largest number of results can be found for the first listed reaction, i.e. d+(t$\mu^-$), which is one of the most important
three-particle reactions in the $\mu$CF cycle in cold liquid hydrogen. We obtained very good agreement
with the experimental data and with some theoretical results, except in the case of work \cite{cohen91}. 
A very good agreement is also obtained for the other two
reactions: for t+(p$\mu^-$) and d+(p$\mu^-$). These results show that the few-body method of the two coupled FH-type equation
(\ref{eq:fh1})-(\ref{eq:fh2}) works extremely well together with the close coupling
expansion method (\ref{eq:expan}). Only three therms in the expansion (\ref{eq:expan}),
i.e. $2\times$(1s+2s+2p) approximation, can provide such
good agreement with the experiments for all three muonic transfer reactions. As we already mentioned
the sign "$2\times$" means that the three term expansion is used within two different expansion functions sets.
Figs. 5 and 6 show our cross sections for t+(p$\mu^-$) and d+(p$\mu^-$) collisions. Again, these results 
are obtained within the different close coupling expansion approximations: 
$2\times$1s, $2\times$(1s+ 2s), and $2\times$(1s+2s+2p). In the first case only two expansion functions 
are used, in the second case only four, and in the third case six expansion functions are used. One can see how significantly the
2p-state contributes to the total muon transfer cross section
when decreasing the collision energy, especially this is seen in the case of d+(p$\mu^-$). A comparison with the
experimental data and other theoretical results in Table I for all three muonic transfer reactions demonstrates that the FH-type
equations and $2\times$(1s+2s+2p) approximation are able to provide reliable results for three-body charge transfer reactions
at low energies. It is also important to mention here, that the pure quantum-mechanical behaviour of the transfer cross
section at low energies, specifically $\sigma_{tr}(\varepsilon_{coll}\sim 0)\sim 1/v_{c.m.}$, has been obtained in this 
calculation. This allowed us to compute the muon transfer rates (\ref{eq:rate7}),  i.e. $\lambda_{tr}(T\rightarrow 0)\approx$const,
and compare these results with the experiments.

Next, the three-body reaction of the atomic $\overline{\rm{H}}$ formation, i.e. reaction (\ref{eq:Pse}) is considered.
We are primarily interested in low energy collisions. Because this is not a muonic system one needs to
switch from muonic to atomic units. 
Also, in the computer program one needs
to change the masses of the particles. Fig. 7 shows our results for the reaction (\ref{eq:Pse}) total cross section in the
framework of different close-coupling approximations in the Eq. (\ref{eq:expan}). One can see, that
the contribution of the 2p-states in each target
become larger while the collision energy becomes smaller.
As in the muonic transfer reactions we found that the cross section of the 
$\overline{\rm{H}}$ formation $\sigma_{\overline{\rm H}}\rightarrow \infty$ as $\varepsilon_{coll}\rightarrow 0$,
i.e. $\sigma_{\overline{\rm H}} v_{c.m.}\approx$const. 
This fact allows us to compute the low energy rate of the $\overline{\rm{H}}$ production. 
For example, one can follow the logic of work \cite{humber87} and estimate
the $\overline{\rm{H}}$ production rate by using the following formula:
$R_{\overline{\rm{H}}}= \sigma_{\overline{\rm H}}N_{\overline{\rm{p}}} l I$. According to \cite{humber87}
$N_{\overline{\rm{p}}}$ is the density of slow antiprotons, 
$I$ is the number of ${\overline{\rm p}}$ traversing the interaction region each second, and
$l$ is the linear dimension of the interaction region. In \cite{humber87} the last parameter was taken as $l=1$ cm.
The product of $l I$ has the unit of velocity, thus
it should be possible to represent the rate $R_{\overline{\rm{H}}}$ as well as the expression (\ref{eq:rate7}), i.e.:
\begin{equation}
R_{\overline{\rm{H}}} = \sigma_{\overline{\rm H}}\ v_{c.m.} N_{\overline{\rm{p}}},
\end{equation}
where $v_{c.m.}$ is the c.m. velocity between ${\overline{\rm p}}$ and the positronium atom Ps.
Our results for ${\overline{\rm{H}}}$ are shown in Table II together with the results for the $\overline{\rm{H}}_{\mu}$
formation reaction (\ref{eq:Psmu}).
At low energies $R_{\overline{\rm{H}}}$ starts taking a constant value as was the case in our previous calculation of
the $\mu^{-}$ transfer reactions. It would be interesting to estimate
the $N_{\overline{\rm{p}}}$ parameter. For example, if we accept
the recent data from the Evaporative Cooling (EC) experiment at the Antiproton Decelerator (AD) at CERN
\cite{andresen2010}: $n_{EC}\approx$4000 very cold ${\overline{\rm{p}}}$ at $T_{EC} \approx9$ K, we would need to
place this quantity of antiprotons in a limited space with a volume $V^{EC}_ {{\overline{\rm{p}}}}$.
The temperature $T_{EC}$ corresponds to the low energy collisions considered in this work: $\sim 10^{-4}$ eV.
In order to obtain the rate $R_{\overline{\rm{H}}}\gtrsim 1$, one would need the following volume:
$V^{EC}_ {{\overline{\rm{p}}}} \approx 4000\sigma_{\overline{\rm H}}\ v_{c.m.}=4000 \times 6.68
\times 10^{-9}\approx 27 \times 10^{-6}$ cm$^3$, where the value for $\sigma_{\overline{\rm H}}\ v_{c.m.}$ is taken
from Table II. If we suppose that the interaction region between ${\overline{\rm{p}}}$ and the Ps atoms
has a cylindrical shape with the length $l_0=1$ cm \cite{humber87}, its radius $r_0$ should be:
$r_0\sim 8.6 \times 10^{-3}$ cm. As one can see $r_0$ has a very small value, although it seems to
us that it still would be possible to adopt this value in some experiments. However, a different situation
arises if we adopt the results of a newer experiment on Adiabatic Cooling (AC) of antiprotons
\cite{gabrielse2011}. The authors of this work obtained $3\times 10^6$ cold antiprotons at temperature 3.5 K!
In this case one would need the following volume:
$V^{AC}_ {{\overline{\rm{p}}}} \approx 3\times 10^6
\sigma_{\overline{\rm H}}\ v_{c.m.}= 3\times 10^6 \times 6.68 \times 10^{-9}
\approx 20 \times 10^{-3}$ cm$^3$. It is easy to compute that in this case the radius is $r_0 \sim$ 0.1 cm.
Finally,  Fig. 8 shows our results for the $\overline{\rm{H}}_{\mu}$
formation cross section. It is clear that in the process (\ref{eq:Psmu}) the contribution of the 2p-states from each target
is becoming even more significant while the collision energy becomes smaller.
Additionally, for the process (\ref{eq:Psmu}) we also compute the numerical value of the quantity:
$\sigma_{\overline{\rm H}_{\mu}}(\varepsilon_{coll}\rightarrow 0)v_{c.m.}\approx$ const.
Table II includes our data for this important parameter together with the $\overline{\rm{H}}_{\mu}$ formation
total cross section. All these results are obtained in the framework of the 2$\times$(1s+2s+2p) close coupling
approximation. These data can be useful in future developments of low energy collision experiments
with participation of cold antiprotons and true muonium atoms.
Next, because of the complexity of the few-body method,
in this work only the total orbital momentum $L=0$ has been taken into account. It was adequate in the case of
slow and ultraslow collisions discussed above.
However, to take into account the important contribution of higher $L$'s at higher collision energies it would be
possible to use Takayanig's Modified Wave Number Approximation (MWNA) method \cite{taka65}.
In the recent work \cite{myfbs2013} the MWNA method has been
successfully applied to a few-body charge transfer reaction.
In conclusion, it is feasible to expect that the FH-type equation formalism
(\ref{eq:fh1})-(\ref{eq:fh2}) could also be an effective tool for computation of the quite
intriguing three-body reaction (\ref{eq:Psmu2}).
This is another process of the $\overline{\rm{H}}_{\mu}$ atom production with a heavy charge transfer from one
center to another \cite{cohen04,annual2010}.
It would be interesting to compare the reaction rates of both processes 
(\ref{eq:Psmu}) and (\ref{eq:Psmu2}) \cite{nagam03}. An additional point to emphasize would be that
in some sense the reaction (\ref{eq:Psmu2}) and the few-body
protonium (Pn) formation reaction \cite{sakimoto02,esry03,macek05,tong06,igarashi2008} show close similarities.
Therefore, it would be good to try to apply the FH-type equation method to the Pn formation problem too.
Also, it seems quite possible to expand in some way or another
the FH-type few-body equation approach and the modified close-coupling expansion method,
Eqs. (\ref{eq:lexpan}) and (\ref{eq:expan}), to very important but challenging low-energy four-particle
rearrangement scattering collisions with the pure Coulomb interaction between the particles, such as 
$\overline{\rm{H}}$+H$\rightarrow (\overline{\rm{p}} \rm{p}^+)_{nl} + (e^+e^-)_{n'l'}$
or, for example,
$\overline{\rm{H}}_\mu$+H$_{\mu}$ $\rightarrow (\overline{\rm{p}} \rm{p}^+)_{nl} + (\mu^+\mu^-)_{n'l'}$.



\section{Appendix} \label{sec:appendix}
The details of the derivation of the angular integrals $S_{\alpha \alpha'}^{ii'}(\rho_j, \rho_k)$ (\ref{eq:final}) are explained below
in this section. 
The configuration triangle, $\bigtriangleup$(123), is determined by the Jacobi vectors $({\vec r_{j3}, \vec \rho_k})$
and should be considered in an arbitrary coordinate system $OXYZ$.
In this initial system the angle variables of the three-body Jacobi vectors $\{\vec r_{j3},\vec \rho_k\}$
have the following values: $\hat r_{j3} = (\theta_j, \phi_j),\ \hat \rho_{k} = (\Theta_k, \Phi_k),\ j \ne k = 1, 2$.
Let us adopt a new coordinate system $O'X'Y'Z'$ in which the axis $O'Z'$ is directed over
the vector $\vec \rho_k$, $\bigtriangleup$(123) belongs to the plain $O'X'Z'$ and
the vertex $k=1$ of $\bigtriangleup$(123) coincides with the origin $O'$ of the new $OX'Y'Z'$.
Fig. 9 shows the specific configuration of $\bigtriangleup$(123) and the new adopted $O'X'Y'Z'$ system. One can see, that
the new angle variables of the Jacobi vectors in the $O'X'Y'Z'$ system have now the following values:
$\hat r'_{j3} = (\nu_k, \pi),\ \hat \rho'_{k} = (0, 0),\ \hat r'_{k3} = (\eta_k, \pi),\ \hat \rho'_{j} = (\omega, \pi)$, here $k=1$ and $j=2$.
The spatial rotational transformation from $OXYZ$ to $O'X'Y'Z'$
has been done with the use of the following Euler angles $(\Phi_k,\Theta_k, \varepsilon)$ \cite{varshal}. 
Taking into account the transformation rule for the bipolar harmonics between new and old coordinate systems,
one can write down the following relationships \cite{varshal}:
\begin{eqnarray}\label{eq:y2}
\left \{ Y_{\lambda}(\hat \rho_k) \otimes Y_l(\hat r_{j3}) \right \}_{LM}^* &=& 
\sum_{m} (D_{Mm}^{L}(\Phi_k, \Theta_k, \varepsilon))^*
\left \{ Y_{\lambda}(\hat \rho_k') \otimes Y_l(\hat r_{j3}')
\right \}^*_{Lm}\\
\left \{ Y_{\lambda'}(\hat \rho_j) \otimes Y_{l'}(\hat r_{k3}) \right \}_{LM}  &=&
\sum_{m'} D_{Mm'}^{L}(\Phi_k, \Theta_k, \varepsilon)
\left \{ Y_{\lambda'}(\hat \rho_j') \otimes Y_{l'}(\hat r_{k3}')
\right \}_{Lm'},
\end{eqnarray}
where $D_{Mm}^{L}(\Phi_k, \Theta_k, \varepsilon)$ are the Wigner functions \cite{varshal}.
The fourfold multiple angular integration $\int d \hat \rho_j \int d \hat \rho_k$
in Eq. (\ref{eq:ss}) can be written in the new variables and be symbolically represented as
$
\int d \hat \rho_j \int d \hat \rho_k = \int_{0}^{\pi} d \omega \sin \omega
\int_{0}^{2\pi} d \varepsilon \int_{0}^{2\pi} d\Phi_k
\int_{0}^{\pi} \sin \Theta_k d\Theta_k.
$
Next, taking into account the normalizing condition for the Wigner functions \cite{varshal}:
\begin{equation}
\int_{0}^{2\pi} d \varepsilon \int_{0}^{2\pi} d\Phi_k
\int_{0}^{\pi} \sin \Theta_k d\Theta_k
(D_{Mm}^L(\Phi_k, \Theta_k, \varepsilon))^*
D_{Mm'}^L(\Phi_k, \Theta_k, \varepsilon)) =
\frac{8 \pi^2}{2L + 1}\delta_{mm'}
\end{equation}
one can obtain the following intermediate expression:
\begin{eqnarray}
S_{\alpha \alpha \prime}^{ii'}(\rho_j, \rho_k) = 2\rho_j \rho_k
\sum_m \frac{8 \pi^2}{2L + 1}
\int_{0}^{\pi} d \omega \sin \omega R_{nl}^i(r_{j3})
\left \{ Y_{\lambda}(0, 0) \otimes Y_l(\hat r'_{j3}) \right \}_{Lm}^*
\nonumber \\
(V_{j3} + V_{jk}) \left \{ Y_{\lambda^\prime}(\hat \rho'_j)
\otimes Y_{l'}(\hat r'_{k3})
\right \}_{Lm'} R_{n'l'}^{i'} (r_{k3})\;.
\label{eq:ssapp}
\end{eqnarray}
Now, let us make the next transformation of $\bigtriangleup$(123) in which the
vertex $j=2$ of $\bigtriangleup$(123) coincides with the centre $O'$
of the $O'X'Y'Z'$ and $O'XYZ$, however the axis $O'Z''$ is directed along $\vec \rho_j$
and $\bigtriangleup$(123) belongs to the plain $O'X''Z''$. This transformation, which converts
the coordinate frame $O'X'Y'Z'$ into $O'X''Y''Z''$ is characterized by
the following Euler angles $(0, \omega, 0)$. Therefore the
vectors $(\vec r_{k3}, \vec \rho_j)$ have the following new variables:
$\hat r^{\prime \prime}_{k3} = (\nu_j, \pi),\ \hat \rho^{\prime \prime}_{j} = (0, 0)$. As a result of this
rotation one can write down the following relationship:
\begin{eqnarray}
\left \{ Y_{\lambda'}(\hat \rho'_j) \otimes Y_{l'}(\hat r'_{k3})
\right \}_{Lm} = \sum_{m'} D_{Mm'}^{L}(0, \omega, 0)
\left \{ Y_{\lambda'}(\hat \rho_j'') \otimes Y_{l'}(\hat r_{k3}'')
\right \}_{Lm'}\;
\label{eq:y3}
\end{eqnarray}
and obtain the following result:
\begin{eqnarray}
S_{\alpha \alpha \prime}^{ii'}(\rho_j, \rho_k) = 2\rho_j \rho_k
\sum_{mm'} \frac{8 \pi^2}{2L + 1}
\int d \omega \sin \omega R_{nl}^i(r_{j3})
\left \{ Y_{\lambda}(0, 0) \otimes Y_l(\hat r'_{j3}) \right \}_{Lm}^*
(V_{j3} + V_{jk})\nonumber \\
%
D_{mm'}^{L}(0, \omega, 0)
\left \{ Y_{\lambda^\prime}(0, 0) \otimes Y_{l'}(\hat r''_{k3})
\right \}_{Lm'} R_{n'l'}^{i'} (r_{k3}).  
\label{eq:soul}
\end{eqnarray}
Now by taking into account that $Y_{lm}(0, 0) = \delta_{m, 0} \sqrt {(2l + 1)/4\pi}$ \cite{varshal},
the bipolar harmonics in (\ref{eq:soul}) are:
\begin{eqnarray}
\left \{ Y_{\lambda}(0, 0) \otimes Y_{l}(\nu_k, \pi)
\right \}_{Lm}^* &=& \sqrt{\frac{2\lambda+1}{4\pi}}  
C_{\lambda 0lm}^{Lm} Y^*_{lm}(\nu_k, \pi),\\
\left \{ Y_{\lambda^\prime}(0, 0) \otimes Y_{l'}(\nu_j, \pi)
\right \}_{Lm'} &=& \sqrt{\frac{2\lambda'+1}{4\pi}}    
C_{\lambda' 0l'm'}^{Lm'} Y_{l'm'}(\nu_j, \pi),
\end{eqnarray}
with the use of these relationships we finally get the convenient for numerical computations
Eq. (\ref{eq:final}). 


\clearpage
\begin{turnpage}
\squeezetable
\begin{table}
\label{tab1}
\caption{Cross sections $\sigma_{tr}$ and rates $\lambda_{tr}$, Eq. (\ref{eq:rate7}), for $\mu^-$ transfer reactions 
from a light hydrogen isotope to a heavier hydrogen isotope at low collision energies together with other theoretical
results and experimental data. The result for unitarity ratio $K_{21}/K_{12}$, Eq. (\ref{eq:unitarity}), 
are also presented for t+(d$\mu)_{1s}$.}
\vspace{2mm}
\centering
\begin{ruledtabular}
\begin{tabular}{lcllllllllll}
Energy, eV & Method &
\multicolumn{3}{c}{t+(d$\mu)_{1s}\rightarrow$ (t$\mu)_{1s}$+d} & \multicolumn{4}{c}{t+(p$\mu)_{1s}\rightarrow$ (t$\mu)_{1s}$+p} &
\multicolumn{2}{c}{d+(p$\mu)_{1s}\rightarrow$ (d$\mu)_{1s}+$p}\\
&  &   $\sigma_{tr} / 10^{-20},\  $cm$^2$ & $\lambda_{tr} / 10^{8}$,\ s$^{-1}$ & $K_{21}/K_{12}$ &
&        $\sigma_{tr} / 10^{-20},\ $cm$^2$ & $\lambda_{tr} / 10^{8}$,\ s$^{-1}$ &
&        $\sigma_{tr} / 10^{-20},\ $cm$^2$ & $\lambda_{tr} / 10^{8}$,\ s$^{-1}$ &\\
\hline
0.001  & FH-type Eqs.:  & 15.4                          & 2.6         & 0.99 &
                & 315.0                          & 65.1          &
                & 663.4                          & 146.0        &\\
               & \cite{melezhik} & 15.8                      & 2.7         &           &
               & 384.4                           & 80.0          &
               & 828.7                           & 170.0        &\\
              & \cite{cohen91} & 21.5   & 3.5            &            &
              & 265.0                            & 55.0          &
              & 650.0                            & 140.0        &\\
              & \cite{kami93}    &18.0   & 2.8           &            &
              &                                      &                 &
              &                                      &                 &\\
              & \cite{igarashi94} & 14.2 &                &           &
              &                                  &              &
              &                                  &              &\\
         & Experiments:  &             & 2.8$\pm$0.5\cite{19}&     &
              &                                  & 58.6$\pm$10\cite{16}&
              &                                  & 84$\pm$13\cite{15}   &\\
         &                        &              & 2.8$\pm$0.3\cite{18}&     &
              &                                  &                                   &
              &                                  & 143$\pm$13\cite{14} &\\
         &                        &             & 2.9$\pm$0.4\cite{17}&     &
              &                                  &                                  &
              &                                  & 95$\pm$34\cite{13}  &\\
\hline
0.01 & FH-type Eqs.:    &  4.84                     & 2.6         & 0.99 &
              & 99.1                          & 64.7         &
              & 208.1                        & 144.8       &\\
         & \cite{melezhik}     & 5.64   &              &           &
              & 128.0                        &              &
              & 283.7                        &              &\\
         & \cite{cohen91}     & 4.8     &              &            &
              & 60.0                          &              &
              & 140.0                        &              &\\
         & \cite{kami93}     &5.0        &               &            &
              &                                 &               &
              &                                 &               &\\
         & \cite{igarashi94}     & 4.44&               &           &
              &                                 &               &
              &                                 &              &\\
\hline
0.04 & FH-type Eqs.:      &  2.37                   & 2.5         & 0.99 &
              & 49.1                          & 64.2         &
              & 103.1                        & 143.4       &\\
        & \cite{melezhik}      & 2.94   &              &           &
              & 63.6                          &              &
              & 140.7                        &              &\\
        & \cite{cohen91}      & 3.1     &              &           &
              & 40.0                          &              &
              & 91.0                          &              &\\
         & \cite{kami93}     &2.5      &               &            &
              &                                &               &
              &                                &               &\\
\hline
0.1   & FH-type Eqs.:     & 1.42                          & 2.4         & 0.99 &
              & 30.7                          & 63.5         &
              & 64.6                          & 142.1       &\\
          &  \cite{melezhik}    & 2.0  &             &          &
              & 39.9                        &             &
              & 87.4                        &             &\\
          & \cite{igarashi94}    &1.35 &              &         &
              &                                 &              &
              &                                 &              &\\
\end{tabular}
\end{ruledtabular}
\end{table}
\end{turnpage}

\clearpage
\begin{table}\label{tab2}
\caption{The total cross sections $\sigma_{\overline{\rm H}}$ and 
$\sigma_{\overline{\rm H}_{\mu}}$ for the reactions (\ref{eq:Pse}) and  (\ref{eq:Psmu}) respectively.
The product of these cross sections and the corresponding center-of-mass velocities $v_{c.m.}$
between $\overline{\rm{p}}$ and Ps=$(e^+ e^-)$, i.e. $\sigma_{\overline{\rm H}} v_{c.m.}$
and between $\overline{\rm{p}}$ and the true muonium atom Ps$_{\mu}=(\mu^+ \mu^-)$, i.e. 
$\sigma_{\overline{\rm H}_{\mu}} v_{c.m.}$ are presented.}
\vspace{2mm}
\centering
\begin{ruledtabular}
\begin{tabular}{lcccccc}
&\multicolumn{2}{c}{$\overline{\rm p}+(e^+ e^-)_{1s}\rightarrow\overline{\rm{H}} + e^-$} &
&\multicolumn{2}{c}{$\overline{\rm p}+(\mu^+\mu^-)_{1s}\rightarrow\overline{\rm{H}}_{\mu} + \mu^-$}\\
\hline
$E$, eV & $\sigma_{\overline{\rm H}}$,\ cm$^2$
               & $\sigma_{\overline{\rm H}}  v_{c.m.}$,\ cm$^3$/s &
               & $\sigma_{\overline{\rm H}_{\mu}}$,\ cm$^2$
               & $\sigma_{\overline{\rm H}_{\mu}}  v_{c.m.}$,\ cm$^3$/s\\
\hline
1.0e-06 & 0.16e-12 & 0.67e-08 &      &                  &               \\
1.0e-05 & 0.50e-13 & 0.67e-08 &      &                  &               \\
1.0e-04 & 0.16e-13 & 0.67e-08 &      &  0.18e-16 & 0.60e-12\\ 
1.0e-03 & 0.50e-14 & 0.66e-08 &      &  0.58e-17 & 0.60e-12\\ 
1.0e-02 & 0.15e-14 & 0.63e-08 &      &  0.18e-17 & 0.59e-12\\ 
5.0e-02 & 0.60e-15 & 0.56e-08 &      &  0.82e-18 & 0.59e-12\\
1.0e-01 & 0.42e-15 & 0.55e-08 &      &  0.58e-18 & 0.59e-12\\
5.0e-01 &                &                 &      &  0.27e-18 & 0.62e-12\\
1.0e-00 &                &                 &      &  0.23e-18 & 0.73e-12\\
\end{tabular}
\end{ruledtabular}
\end{table}


\clearpage

%
\clearpage
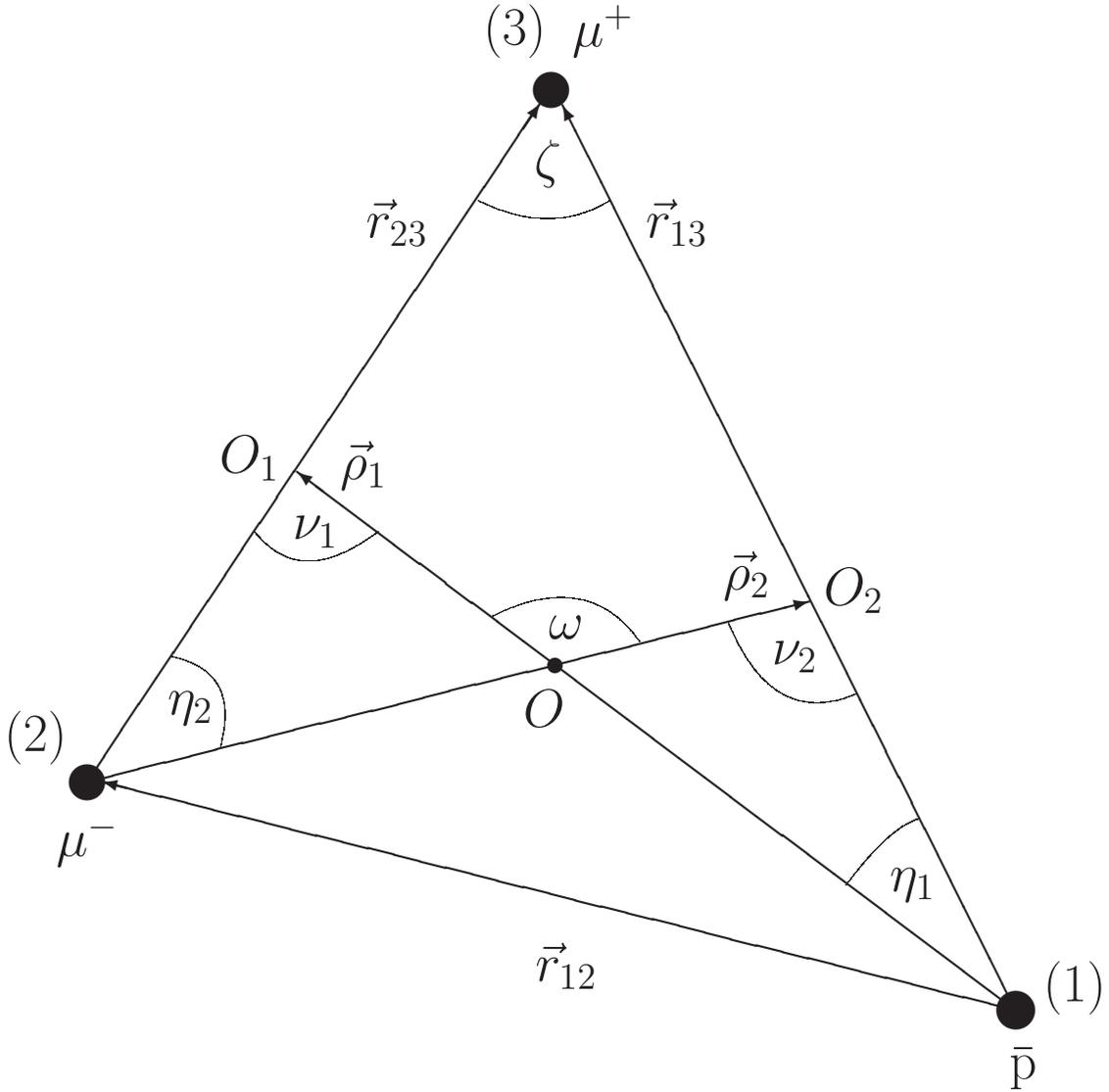
\begin{figure}
%

\begin{center}
{
\unitlength 1.00mm

\begin{picture}(120,190) (10,-20)

\thicklines
\put(5,51){\circle*{5}}      
\put(68,145){\circle*{5}}   
\put(131,20){\circle*{12}}    

\put(68.5,66.8){\circle*{2}}   

\put(5,51){\vector(2,3){61.5}}
\put(5,51){\vector(4,1){98}}
\put(131,20){\vector(-4,1){124}}
\put(131,20){\vector(-1,2){61.5}}
\put(131,20){\vector(-4,3){97.5}}

\put(67.5,135){\makebox(0,0)[cc]{\LARGE $\zeta$}}

\put(47,127){\makebox(0,0)[cc]{\LARGE $\vec r_{23}$}}
\put(85,127){\makebox(0,0)[cc]{\LARGE $\vec r_{13}$}}
\put(70,26){\makebox(0,0)[cc]{\LARGE $\vec r_{12}$}}

\put(36,85){\makebox(0,0)[cc]{\LARGE $\nu_1$}}
\put(101,68){\makebox(0,0)[cc]{\LARGE $\nu_2$}}

\put(27,95){\makebox(0,0)[cc]{\LARGE $O_{1}$}}
\put(109,77){\makebox(0,0)[cc]{\LARGE $O_{2}$}}
\put(67,61){\makebox(0,0)[cc]{\LARGE $O$}}

\put(42.5,94){\makebox(0,0)[cc]{\LARGE $\vec \rho_1$}}
\put(94.5,79){\makebox(0,0)[cc]{\LARGE $\vec \rho_2$}}

\put(70,72){\makebox(0,0)[cc]{\LARGE $\omega$}}

\put(19,61){\makebox(0,0)[cc]{\LARGE $\eta_2$}}
\put(117,37){\makebox(0,0)[cc]{\LARGE $\eta_1$}}

\put(139,23){\makebox(0,0)[cc]{\LARGE $(1)$}}
\put(132,12){\makebox(0,0)[cc]{\LARGE ${\bar{\rm p}}$}}

\put(-2,57){\makebox(0,0)[cc]{\LARGE $(2)$}}
\put(5,43){\makebox(0,0)[cc]{\LARGE $\mu^-$}}

\put(63,153){\makebox(0,0)[cc]{\LARGE $(3)$}}
\put(75,153){\makebox(0,0)[cc]{\LARGE $\mu^+$}}

\linethickness{0.02mm}
\qbezier(58,130)(68,125)(76,130)  
\qbezier(60,73.5)(73,80)(80,70)  
\qbezier(16.5,68)(26,66)(23,55.5)  
\qbezier(108,37)(114,45)(118,46)  
\qbezier(27.7,85)(34,77)(44.5,85)  
\qbezier(92,72.7)(98,58)(109.5,63)  

\end{picture}
\vspace{-2cm}
\caption{The configurational triangle of a three-charged-particle system (123).
$(\overline{{\mbox p}}\  \mu^- \mu^+)$ is presented together with the Jacobi
coordinates, the inter-particle vectors, and the angles between the vectors and coordinates. $O$ is the
center of mass of the few-body system system, $O_1$ and $O_2$ are the center of masses of the
targets $\mu^-\mu^+$ and $\overline{\rm{p}}\mu^+$ respectively.}
\label{fig1}
}\end{center}
\end{figure}

%
%

\clearpage
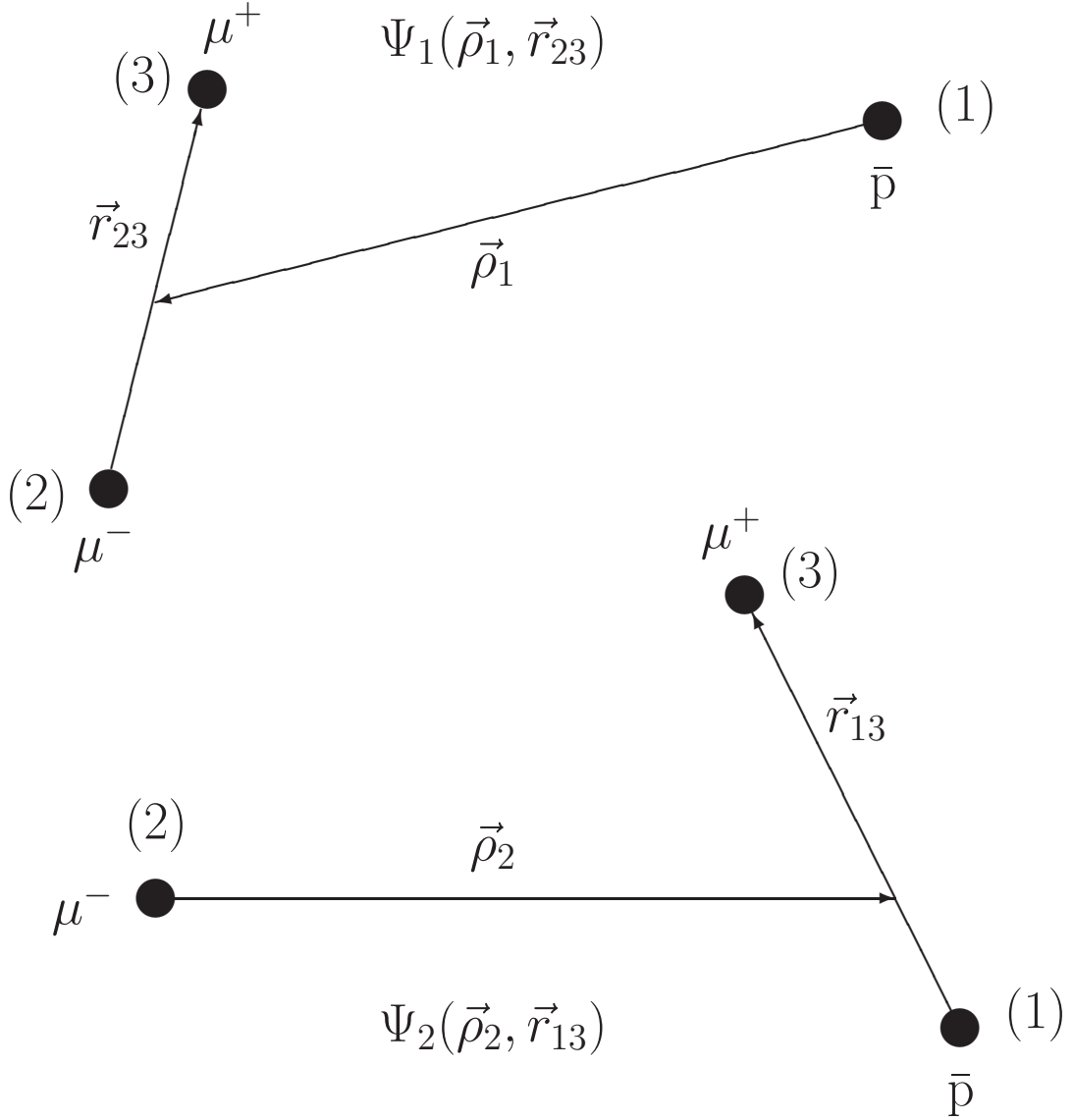
\begin{figure}
\begin{center}
{
\begin{picture}(800,300)(-100,0)

\put(-18,58){\circle*{20}}           
\put(20,212){\circle*{20}}         
\put(280,200){\circle*{20}}        
\put(-58,50){\LARGE {(2)}}      
\put(-32,30){\LARGE {$\mu^-$}}    
\put(-17,211){\LARGE {(3)}}    
\put(18,230){\LARGE {$\mu^+$}}    
\put(300,200){\LARGE {(1)}}    
\put(275,170){\LARGE {$\bar{\rm p}$}}    
\put(-14,160){\makebox(0,0)[cc]{\LARGE $\vec r_{23}$}}
\put(130,145){\makebox(0,0)[cc]{\LARGE $\vec \rho_1$}}
\put(130,230){\makebox(0,0)[cc]{\LARGE $\Psi_1(\vec \rho_1, \vec r_{23})$}}
\thicklines
\put(-17,66){\vector(1,4){34.5}}          
\put(280,200){\vector(-4,-1){280}}      

\put(0,-100){\vector(1,0){285}}    
\put(0,-100){\circle*{20}}           
\put(310,-150){\circle*{20}}           
\put(227,17){\circle*{20}}           
\put(310,-150){\vector(-1,2){80}}   
\put(130,-80){\makebox(0,0)[cc]{\LARGE $\vec \rho_2$}}
\put(270,-30){\makebox(0,0)[cc]{\LARGE $\vec r_{13}$}}
\put(-12,-77){\LARGE {(2)}}      
\put(-40,-110){\LARGE {$\mu^-$}}    
\put(327,-150){\LARGE {(1)}}    
\put(305,-180){\LARGE {$\bar{\rm p}$}}    
\put(240,20){\LARGE {(3)}}    
\put(210,35){\LARGE {$\mu^+$}}    
\put(130,-150){\makebox(0,0)[cc]{\LARGE $\Psi_2(\vec \rho_2, \vec r_{13})$}}
\end{picture}
\vspace{6cm}
\caption{Two asymptotic spacial configurations of the 3-body system (123), or more specifically
$(\bar{\mbox{p}}, \mu^-, \mu^+)$ which is considered in this work. The few-body
Jacobi coordinates $(\vec \rho_i, \vec r_{jk})$, where $i\ne j\ne k=1,2,3$  are also shown together with the 3-body wave function
components $\Psi_1$ and $\Psi_2$: $\Psi = \Psi_1 + \Psi_2$ is the total wave function of the 3-body system.}
}\end{center}
\label{fig:fig2}
\end{figure}

%

\clearpage
\begin{figure}            
\begin{center}
\includegraphics*[scale=1.0,width=40pc,height=45pc]{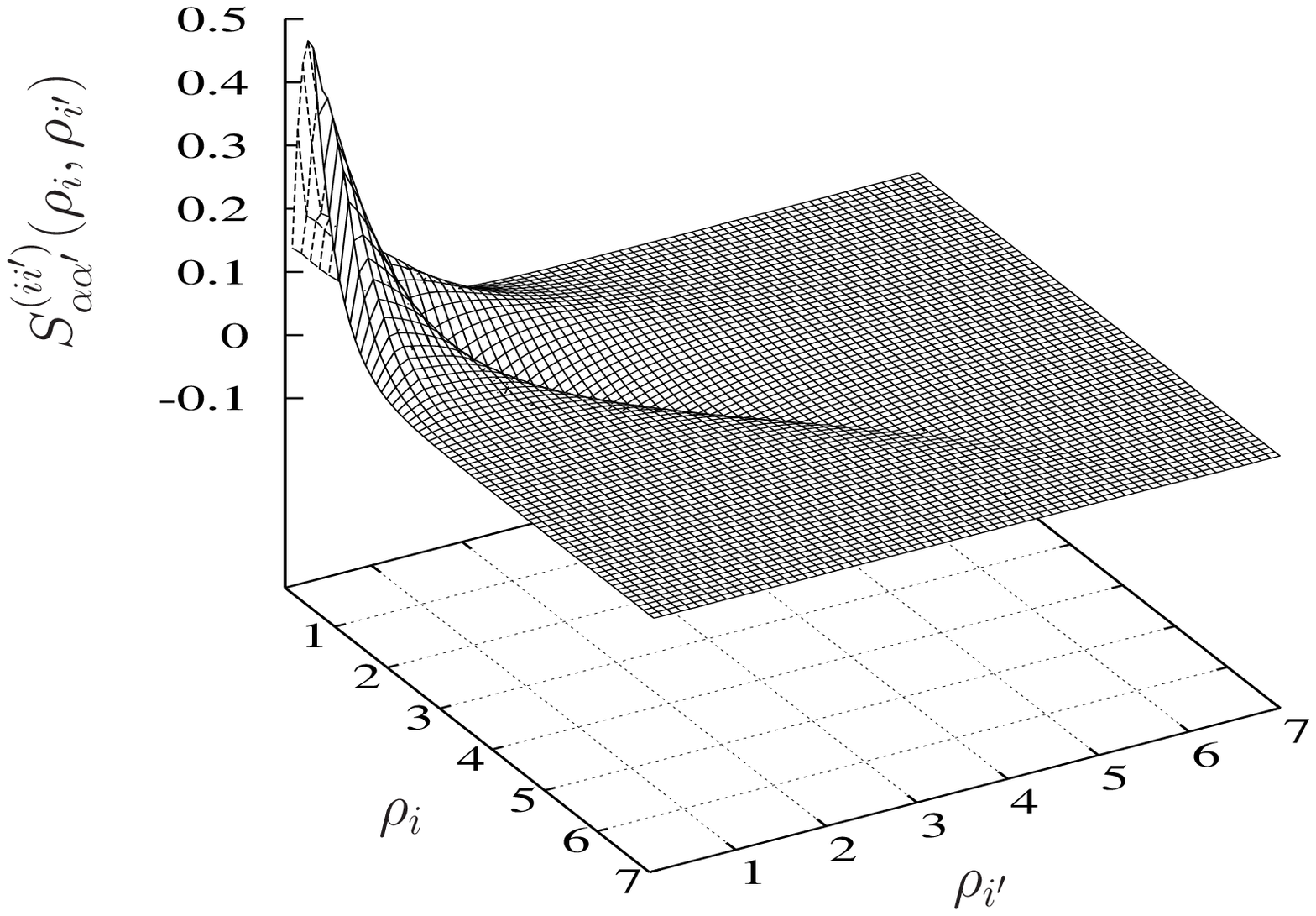}
\caption{The angular integral $S_{\alpha \alpha'}^{ii'}(\rho_i, \rho_{i'})$, Eq. (\ref{eq:omega}),
in the input channel $\overline{\rm p}+(\mu^+\mu^-)$ when $\alpha=$1s and $\alpha'=$1s.}
\label{fig:fig3}
\end{center} 
\end{figure}
\clearpage
\begin{figure}            
\begin{center}
\includegraphics*[scale=1.0,width=40pc,height=45pc]{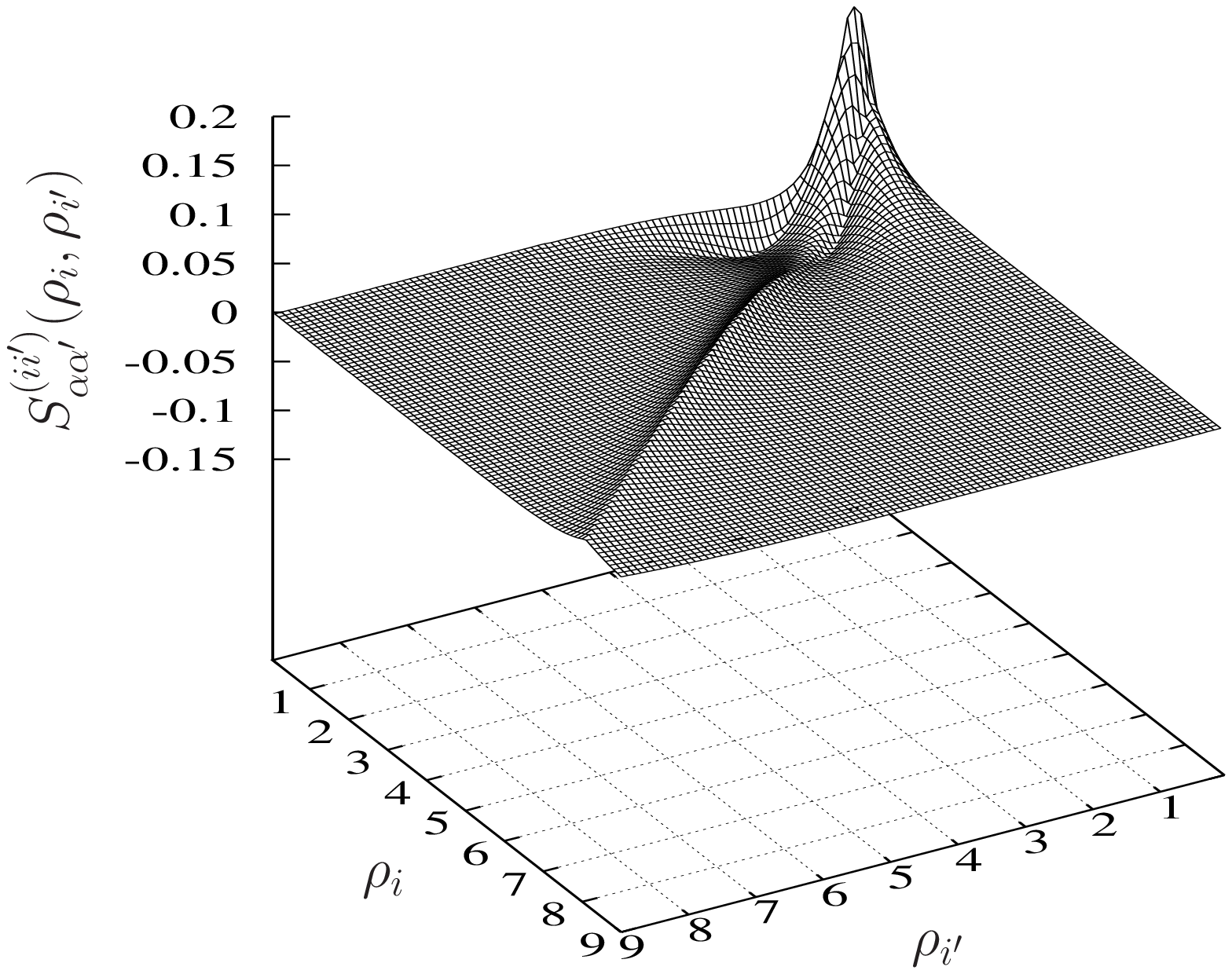}
\caption{The angular integral $S_{\alpha \alpha'}^{ii'}(\rho_i, \rho_{i'})$, Eq. (\ref{eq:omega}),
in the input channel $\overline{\rm p}+(\mu^+\mu^-)$ when $\alpha=$1s and $\alpha'=$2s.}
\label{fig:fig4}
\end{center} 
\end{figure}



%


\clearpage
\begin{figure}            
\begin{center}
\includegraphics*[scale=0.68]{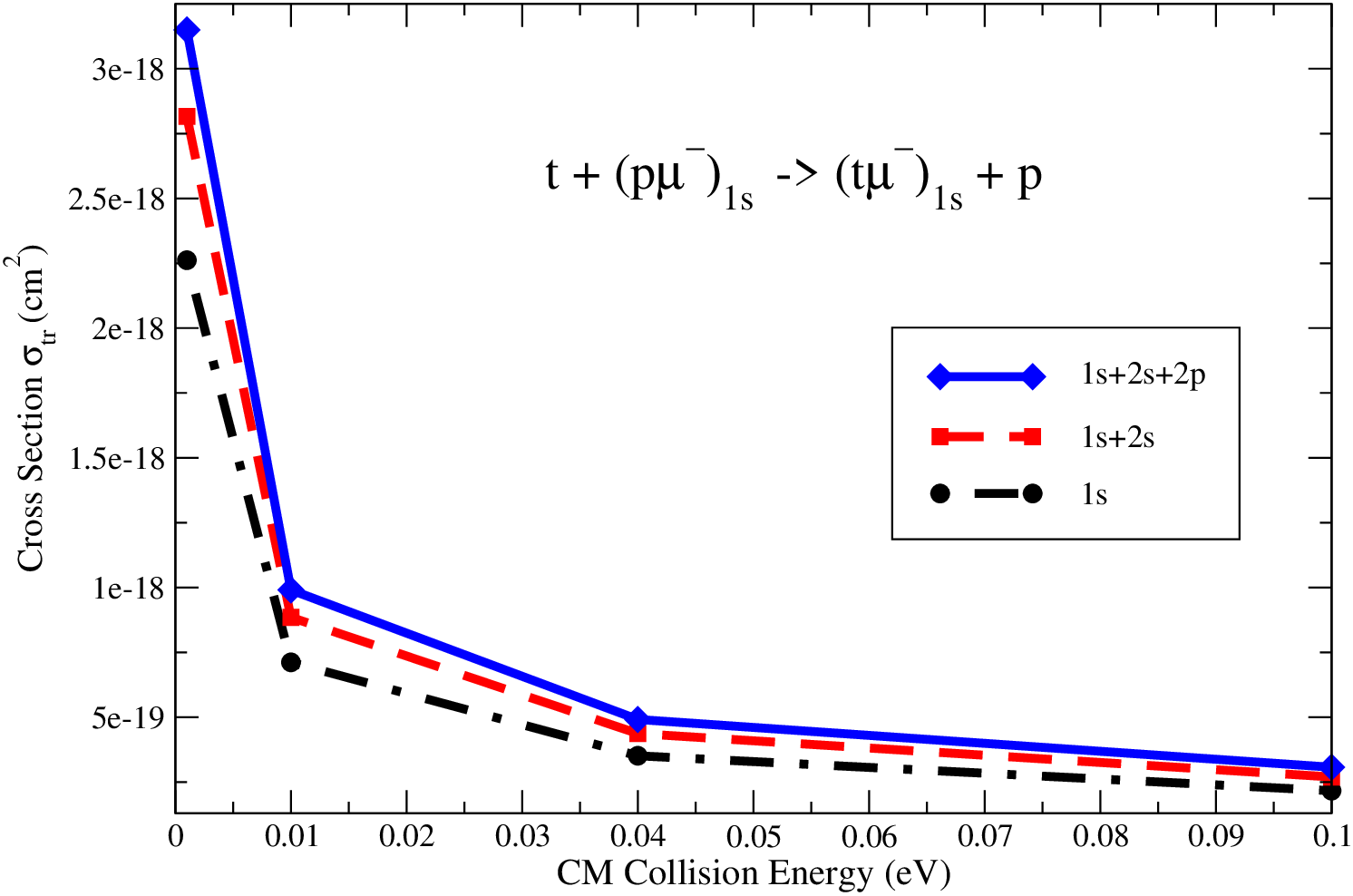}
\caption{Low energy cross sections for a muon transfer reaction in the
t+(p$\mu^-)_{1s}$ collision. 
Results are shown for the two-level
2$\times$1s, four-level 2$\times$(1s+2s), and six-level 2$\times$(1s+2s+2p) close-coupling approximations.
}
\label{fig:fig7} \end{center} \end{figure}

\clearpage
\begin{figure}
\begin{center}            
\includegraphics*[scale=0.68]{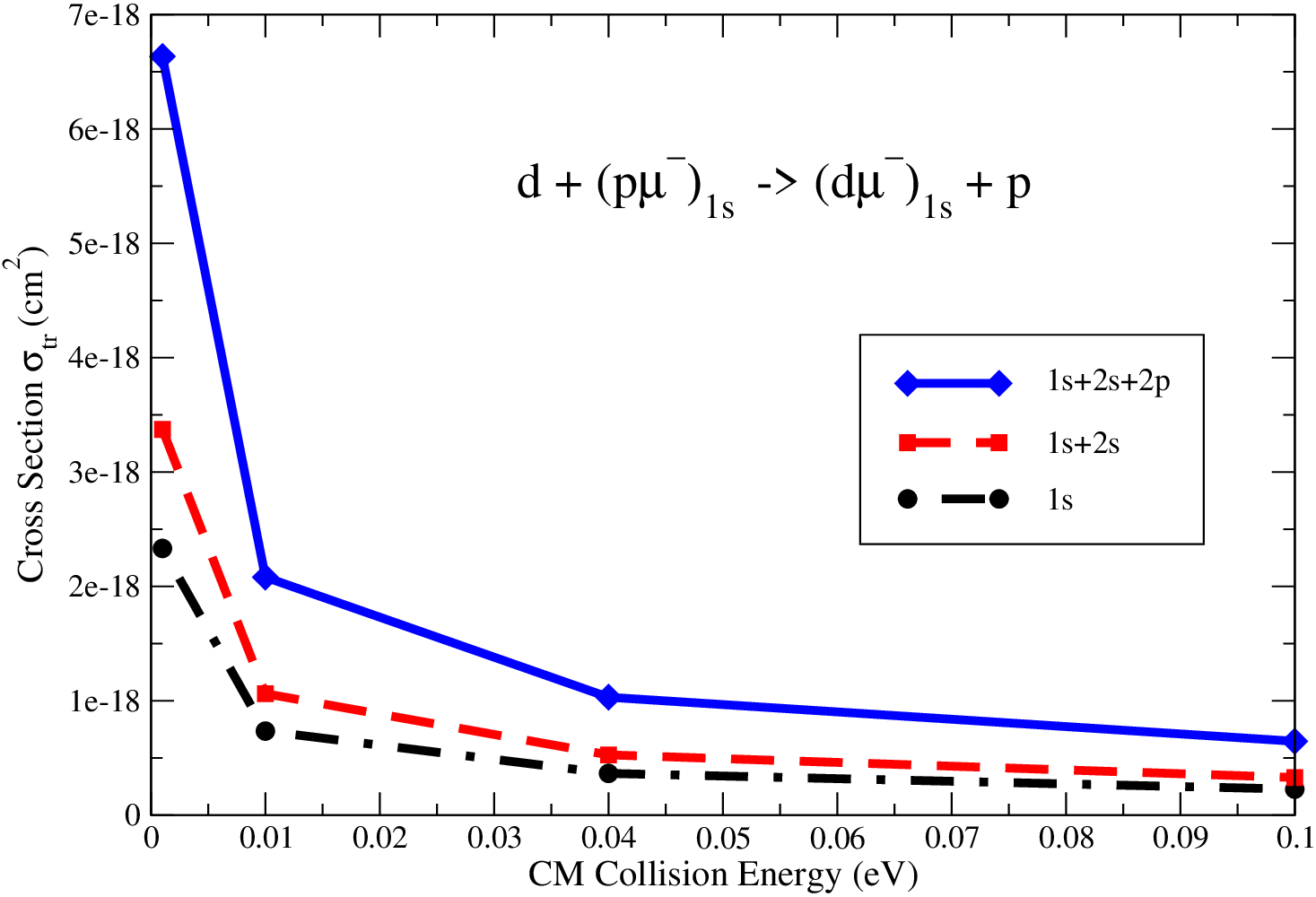}
\caption{Low energy cross sections for a muon transfer reaction in the
d+(p$\mu^-)_{1s}$ collision. 
Results are shown for the two-level
2$\times$1s, four-level 2$\times$(1s+2s), and six-level 2$\times$(1s+2s+2p) close-coupling approximations.}
\label{fig:fig8} \end{center} \end{figure}

\clearpage
\begin{figure}            
\begin{center}
\includegraphics*[scale=0.68]{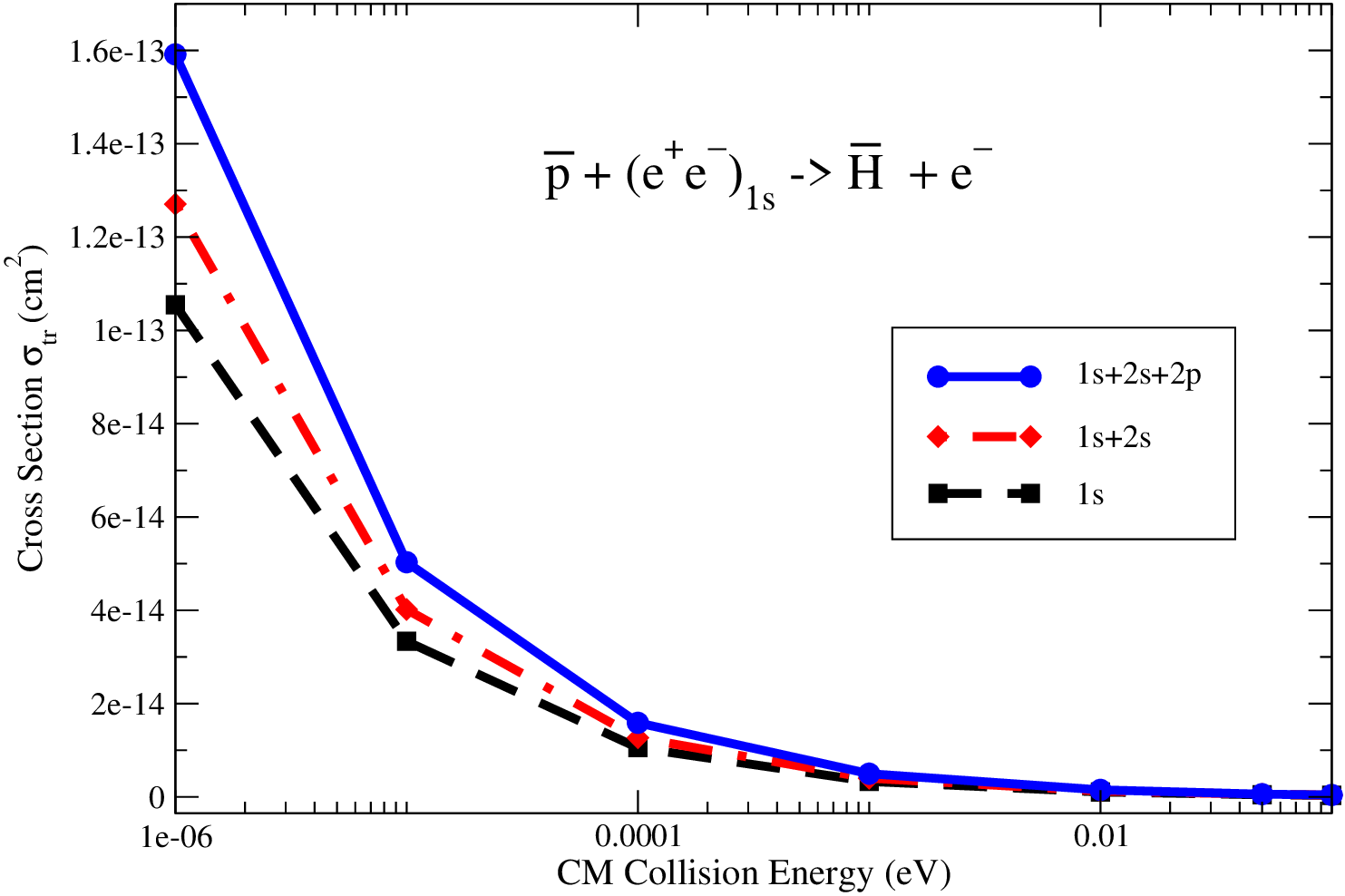}
\caption{Low energy cross sections for an antihydrogen atom formation reaction (\ref{eq:Pse}).
Results are shown for the two-level
2$\times$1s, four-level 2$\times$(1s+2s), and six-level 2$\times$(1s+2s+2p) close-coupling approximations.}
\label{fig:fig9} \end{center} \end{figure}

\clearpage
\begin{figure}            
\begin{center}
\includegraphics*[scale=0.68]{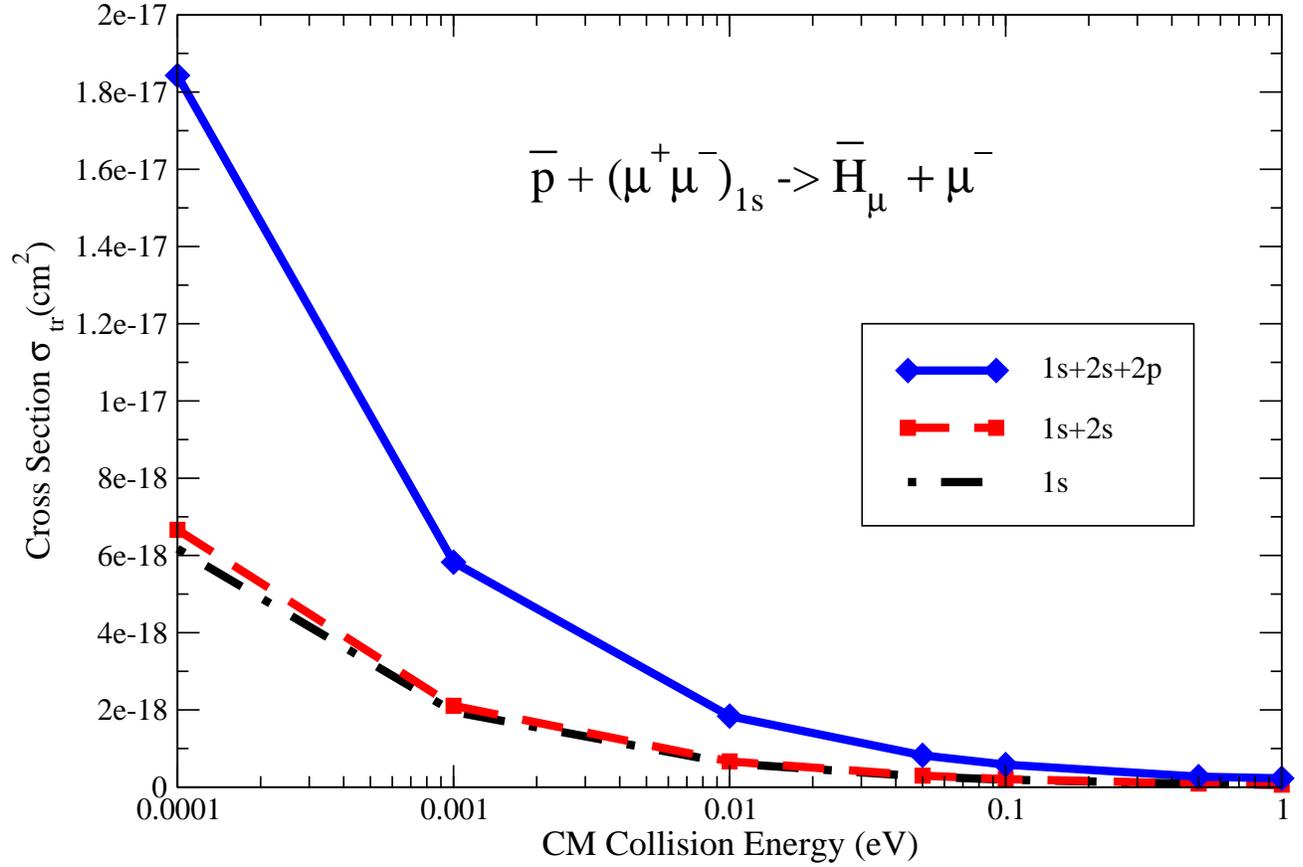}
\caption{Low energy cross sections for a muonic antihydrogen formation reaction (\ref{eq:Psmu}).
Results are shown for the two-level
2$\times$1s, four-level 2$\times$(1s+2s), and six-level 2$\times$(1s+2s+2p) close-coupling approximations.}
\label{fig:fig10} \end{center} \end{figure}

\clearpage
\begin{figure}            
\begin{center}
\includegraphics*[scale=0.680]{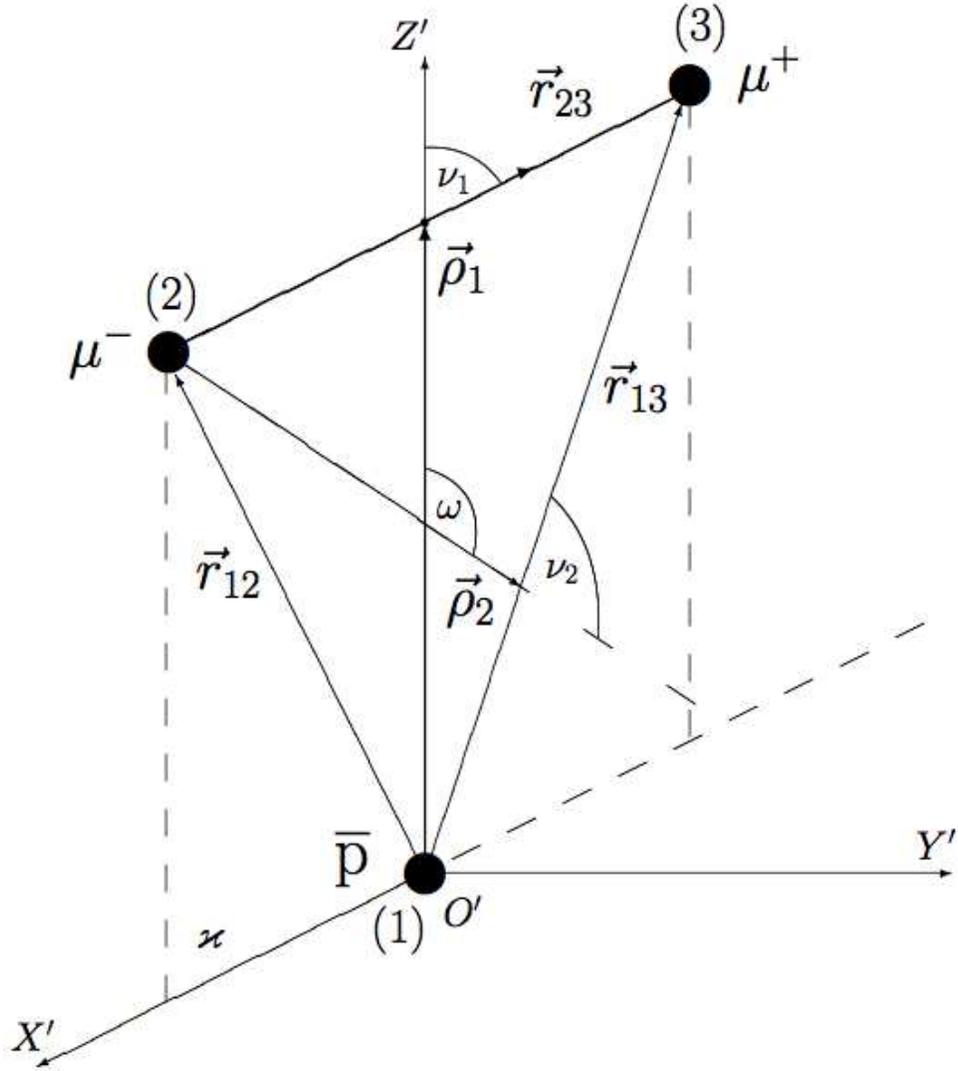}
\caption{   
The configurational $\bigtriangleup$(123) in the case of the $\overline{{\mbox p}}+(\mu^-\mu^+)$ collision
is shown together with the new $O'X'Y'Z'$ cartesian coordinate system after the rotational-translational
transformation from the initial $OXYZ$ system (see Appendix, Sect. IV). $OXYZ$ is not shown here.}
\label{fig:fig11} \end{center} \end{figure}






\end{document}